\documentclass[prd,twocolumn,showpacs,preprintnumbers,amsmath,amssymb]{revtex4}

\usepackage{graphicx}
\usepackage{epsfig}
\usepackage{dcolumn}
\usepackage{bm}

 \newcommand{\zr}[1]{\mbox{\hspace*{#1em}}}
 \newcommand{\ID}{\mbox{{\sf 1}\zr{-0.14}\rule{0.04em}{1.55ex}\zr{0.1}}}
 
 \newcommand{\RR}{\mbox{\zr{0.1}\rule{0.04em}{1.6ex}\zr{-0.05}{\sf R}}}

\begin{document}

\preprint{\sf FAU-TP3-01/11, hep-ph/0202194}

\title{On the Infrared Exponent for Gluon and Ghost\\ 
Propagation in Landau Gauge QCD}

\author{Christoph Lerche}
\altaffiliation[Now at ]{Instituto de F\'\i sica Corpuscular, Edificio
Institutos de  Paterna, Apartado 22085, cp: 46071. Valencia, Spain}
\author{Lorenz von Smekal}%
\affiliation{%
Institut f\"ur Theoretische Physik III, Universit\"at
Erlangen-N\"urnberg, Staudtstr. 7, D-91058 Erlangen, Germany}%

\date{\today}

\begin{abstract}
In the covariant description of confinement, one expects the 
ghost correlations to be infrared enhanced. 
Assuming ghost dominance, the long-range behavior of gluon and ghost
correlations in Landau gauge QCD is determined by one exponent
$\kappa $. The gluon propagator is infrared finite (vanishing)   
for $\kappa =1/2$ ($\kappa > 1/2$) which is still under debate. 
Here, we study critical exponent and coupling 
for the infrared conformal behavior from the asymptotic form of the solutions
to the Dyson-Schwinger equations in an ultraviolet finite expansion scheme. 
The value for $\kappa $ is directly related to the ghost-gluon vertex. 
Assuming that it is regular in the infrared, one obtains $\kappa \simeq 0.595
$. This value maximizes the critical coupling  $\alpha_c(\kappa)$, yielding
$\alpha_c^{max} \simeq (4\pi/N_c)\,  0.709 \approx 2.97 $ for $N_c=3$. 
For larger $\kappa $ the vertex acquires an infrared
singularity in the gluon momentum, smaller ones imply infrared singular ghost
legs. Variations in $\alpha_c$ remain within 5\% from $\kappa = 0.5$ to
$0.7$.  
Above this range, $\alpha_c$ decreases more rapidly 
with $\alpha_c \to 0^+$ as $\kappa \to 1^-$ which sets
the upper bound on $\kappa $.
\end{abstract}

\maketitle

\section{Introduction}
\label{sec:level1}

In gauge theories without Higgs mechanism, particles carrying 
the global charges of the gauge group cannot strictly be localized.
Localized physical states are necessarily neutral in QED and colorless 
in QCD. The extension to {\em all} gauge invariant and thus physical states 
is possible only with a mass gap in the physical world. 
Then, color-electric charge superselection sectors do not arise 
in QCD and one concludes confinement. 

The necessary conditions for this were formulated 
more than twenty years ago. In the next subsection we 
briefly recall these conditions, and how they constrain the infrared 
behavior of ghost and gluon propagators in Landau gauge QCD. 
Based on linear-covariant gauges, their derivation
may not fully be divorced from perturbation theory.   
Their essence is quite generic and summarized in the Kugo-Ojima criterion
which should apply in one way or another, whenever some 
form of BRS cohomology construction does for gauge theories. 
One way towards a non-perturbative definition of the Landau gauge is provided
via stochastic quantization for which the full 5-dimensional BRS machinery is
in the garage.~The time-independent diffusion equation of this formulation
is closely related to the Dyson-Schwinger equations (DSEs) in 4 dimensions as
we describe next. Some of the necessary extensions, which have already been
implemented in previous DSE studies of infrared
exponents for other reasons, imply the Kugo-Ojima criterion. 
We summarize these studies, and how they are confirmed qualitatively  in this
way, at the end of the introduction.    


In Sec.~II, we set up the DSE structures relevant for our present study.   
We summarize general properties of the ghost-gluon vertex, most importantly 
its non-renormalization and ghost-antighost symmetry in Landau gauge,
which will be essential for infrared critical exponents and coupling later.
We then present the ultraviolet subtraction procedure with the 
special care necessary to make sure that it does not artificially 
affect the infrared. Some confusion arose recently concerning the relation
between asymptotic infrared expansions and the renormalization group which we
first clarify in Sec.~III. We then review the non-perturbative 
definition of the running coupling that is based on the non-renormalization 
of the ghost-gluon vertex in Landau gauge, and show that in 4 dimensions 
it approaches a constant $\alpha_c$ in the infrared whenever this vertex has an
asymptotic conformal behavior also. As a byproduct of the vertex
non-renormalization, the infrared behavior of both propagators thereby
results to be determined by one unique exponent $\kappa $ in any given
dimension. The general machinery to determine infrared critical exponent and
coupling is outlined in Sec.~IV. There, we also discuss the results with an
additional regularity assumption on the vertex in the infrared, which in 4
dimensions leads to the values $\kappa \approx 0.595$ and $\alpha_c \equiv
\alpha_c^{max} \simeq (4\pi/N_c)\,  0.709 \approx 2.97 $ for $N_c=3$.  
We furthermore discuss the infrared transversality of the vertex and show
how this resolves an apparent contradiction with a previous study.  

We then discuss more general vertices involving an additional exponent which
controls singularities in its external momenta to discuss bounds on
$\alpha_c$ and $\kappa $. Thereby we will find that 
values of $\kappa $ smaller than that for the regular vertex imply infrared
divergences in ghost legs, whereas larger ones lead to an infrared divergence
of the vertex in the  gluon momentum. 
While the latter can only come together with an infrared
vanishing gluon propagator, which will always over-compensate this
divergence, the former add to the infrared enhancement of ghost
exchanges. In particular, this would have to happen for an infrared finite
gluon propagator (with $ \kappa = 0.5 $) as presently favored by lattice
simulations.      

Our summary and conclusions are given in Sec.~V, and
we include 2 appendices which may provide the interested reader with 
some more technical details.


\subsection{The Kugo-Ojima confinement criterion}
\label{sec:}

Within the framework of BRS algebra, 
completeness of the nilpotent BRS-charge $Q_B$, the  
generator of the BRS transformations, in a state space $\mathcal{V}$
of indefinite metric is assumed. 
The semi-definite {\em physical} subspace 
${\mathcal{V}}_{\mbox{\tiny phys}}  = \mbox{Ker}\, Q_B  $
is defined on the basis of this algebra by those states which are annihilated
by the BRS charge $Q_B$. 
Since $Q_B^2 =0 $, this subspace contains the space $ \mbox{Im}\, Q_B $
of so-called daughter states which are images of others, their parent states
in $\mathcal{V}$. 
A physical Hilbert space is then obtained as (the completion of) the 
covariant space of equivalence classes ${\mbox{Ker}\, Q_B}/{\mbox{Im}\, Q_B}
$, the BRS-cohomology of states in the kernel modulo those in the image of
$Q_B$, which is isomorphic to the space ${\mathcal{V}}_s$ of BRS singlets. 
It is easy to see that the image is furthermore contained in the orthogonal
complement of the kernel (given completeness they are identical).
It follows that states in $\mbox{Im}\, Q_B$
do not contribute to the inner product in $\mathcal{V_{\mbox{\tiny phys}}}$.   

Completeness is thereby important in the proof of positivity for physical
states \cite{Kug79,Nak90}, because it assures the absence of metric
partners of BRS-singlets, so-called ``singlet pairs''.
With completeness, all states in $\mathcal{V}$ are either BRS
singlets in ${\mathcal{V}}_s$ or belong to so-called quartets which are 
metric-partner pairs of BRS-doublets (of parent with daughter states).
And this then exhausts all possibilities. The generalization of the
Gupta-Bleuler condition on physical states, $Q_B |\psi\rangle = 0$ in
$\mathcal{V}_{\mbox{\tiny phys}}$, eliminates half of these metric partners
leaving unpaired states of zero norm 
which do not contribute to any observable.
This essentially is the quartet mechanism: 
Just as in QED, one such quartet, the elementary quartet, is formed by
the massless asymptotic states of longitudinal and timelike gluons together 
with ghosts and antighosts which are thus all unobservable. 
In contrast to QED, however, one expects the quartet mechanism also 
to apply to transverse gluon and quark states, as far as they exist
asymptotically. A violation of positivity for such states then entails
that they have to be unobservable also. 
Increasing evidence for this has been seen in the 
transverse gluon correlations over the last years \cite{Alk01}. 

But that is only one aspect of confinement in this description.   
In particular, asymptotic transverse gluon and quark states
may or may not exist in the indefinite metric space $\mathcal{V}$. If either 
of them do, and the Kugo-Ojima criterion is realized (see below), they
belong to unobservable quartets. Then, the BRS-transformations of their
asymptotic fields entail that they form these quartets together with
ghost-gluon and/or ghost-quark bound states, respectively \cite{Nak90}.
It is furthermore  crucial for confinement, however, to have a mass gap in
transverse gluon correlations. The massless transverse gluon
states of perturbation theory must not exist even though they would
belong to quartets due to color antiscreening and superconvergence in QCD for
less than 10 quark flavors \cite{Oeh80,Nis94,Alk01}. 

Confinement depends on the 
realization of the unfixed global gauge symmetries.
The identification of gauge-invariant physical states, which are   
BRS-singlets, with color singlets is possible only if the charge of global
gauge transformations 
is BRS-exact {\em and} unbroken. The sufficent conditions for this are 
provided by the Kugo-Ojima criterion: Considering the 
globally conserved current     
\begin{equation} 
    J^a_\mu = \partial_\nu F_{\mu\nu}^a  + \{ Q_B , D_{\mu}^{ab} \bar c^b \} 
    \qquad (\mbox{with} \; \partial_\mu J^a_\mu = 0 \,) \; ,
       \label{globG}
\end{equation}
one realizes that the first of its two terms corresponds to a coboundary 
with respect to the space-time exterior derivative while the second term 
is a BRS-coboundary. Denoting their charges by $G^a$ and $N^a$, respectively, 
\begin{equation} 
      Q^a =  \int d^3x \Big(  \partial_i F_{0 i}^a   +  
             \{ Q_B , D_{0}^{ab} \bar c^b \} \Big)  =  G^a + N^a \, .
        \label{globC}
\end{equation}
For the first term herein there are only two options, it is either ill-defined
due to massless states in the spectrum of $\partial_\nu F_{\mu\nu}^a $, or else
it vanishes. 

In QED massless photon states contribute to the analogues of both currents
in~(\ref{globG}), and both charges on the r.h.s. in (\ref{globC}) are
separately ill-defined. One can employ an arbitrariness in the 
definition of the generator of the global gauge transformations
(\ref{globC}), however, to multiply the first term by a suitable
constant so chosen that these massless contributions cancel.
In this way one obtains a well-defined and unbroken global
gauge charge which replaces the naive definition in (\ref{globC})
above \cite{Kug95}. Roughly speaking, there are two independent structures in
the globally conserved gauge currents in QED which both contain massless
photon contributions. These can be combined 
to yield one well-defined charge as the generator of global gauge
transformations leaving any other combination spontaneously broken,
such as the displacement symmetry which led to the identification of
the photon with the massless Goldstone boson of its spontaneous breaking 
\cite{Nak90,Fer71}. 

If $\partial_\nu F_{\mu\nu}^a $ contains no massless
discrete spectrum on the other hand, {\it i.e.}, if there is no massless
particle pole in the Fourier transform of transverse gluon correlations, then
$G^a \equiv 0$.
In particular, this is the case for channels containing massive vector fields
in theories with Higgs mechanism, and it is expected to be also the case in
any color channel for QCD with confinement for which it actually represents one
of the two conditions formulated by Kugo and Ojima. 
In both these situations one first has equally, however, 
\begin{equation}
                       Q^a \, = \, N^a \, = \, \Big\{   Q_B \, , 
       \int d^3x \,   D_{0}^{ab} \bar c^b \Big\} \; ,
\end{equation}
which is BRS-exact. The second of the two conditions for confinement
is that this charge be well-defined in the whole of the indefinite metric space
$\mathcal{V}$. Together these conditions 
are sufficient to establish that all BRS-singlet physical
states are also color singlets, and that all colored states
are thus subject to the quartet mechanism. The 
second condition thereby provides the essential 
difference between Higgs mechanism and confinement. 
The operator $D_\mu^{ab}\bar c^b$ determining the charge $N^a$ will in
general contain a  {\em massless} contribution from the elementary
quartet due to the asymptotic field $\bar\gamma^a(x)$ in the  
antighost field,  $\bar c^a\, \stackrel{x_0 \to \pm\infty}{\longrightarrow}
\, \bar\gamma^a + \cdots $ (in the weak asymptotic limit), 
\begin{equation}
          D_\mu^{ab}\bar c^b \; \stackrel{x_0 \to \pm\infty}{\longrightarrow}
              \;   ( \delta^{ab} + u^{ab} )\,   \partial_\mu \bar\gamma^b(x) +
                 \cdots  \;  .
\end{equation}
Here, the dynamical parameters $ u^{ab} $ determine the contribution 
of the massless asymptotic state to the composite field $g f^{abc} A^c_\mu
\bar c^b  \, \stackrel{x_0 \to \pm\infty}{\longrightarrow}  \,
u^{ab} \partial_\mu \bar\gamma^b + \cdots $. These parameters can be obtained
in the limit $p^2\to 0$ from the Euclidean correlation functions of this
composite field, {\it e.g.},
\begin{eqnarray}
&& \hspace{-1cm} \int d^4x \; e^{ip(x-y)} \,
\langle  \; D^{ae}_\mu c^e(x) \; gf^{bcd}A_\nu^d(y) \bar c^c (y) \; \rangle
\; \equiv   \nonumber \\
&& \hskip 3cm
\Big(\delta_{\mu \nu} -\frac{p_\mu p_\nu}{p^2} \Big) \, u^{ab}(p^2)
\; .  \label{Corru}
\end{eqnarray}
The theorem by Kugo and Ojima asserts that all $Q^a = N^a$ are
well-defined in the whole of  $\mathcal{V}$ (and do not suffer from
spontaneous breakdown), if and only if
\begin{eqnarray}
                 u^{ab} \equiv u^{ab}(0)  \stackrel{!}{=} - \delta^{ab} \; .
\label{KO1}
\end{eqnarray}
Then, the massless states from the elementary quartet do not contribute to 
the spectrum of the current in $N^a$, and the equivalence between physical
BRS-singlet states and color singlets is established \cite{Kug79,Nak90,Kug95}.

In contrast, if $\mbox{det}(  \ID + u ) \not=0$, the global
gauge symmetry generated by the charges $Q^a$ in eq.~(\ref{globC}) is
spontaneuosly broken in each channel in which the gauge potential 
contains an asymptotic massive vector field \cite{Kug79,Nak90}. 
While this massive vector state 
results to be a BRS-singlet, the massless Goldstone boson states which 
usually occur in some components of the Higgs field, replace the 
third component of the vector field in the elementary
quartet and are thus unphysical. 
Since the broken charges
are BRS-exact, this {\em hidden} 
symmetry breaking is not directly observable in the physical Hilbert space.  

The different scenarios are classified 
according to the realization of the global gauge symmetry on the
whole of the indefinite metric space of covariant gauge
theories. If it is unbroken, as in QED and QCD, 
the first condition is crucial for confinement.
Namely, it is then necessary  to
have a mass gap in the transverse gluon correlations, 
since otherwise one could in principle have
{\em non-local} physical (BRS-singlet and thus gauge-invariant) states with
color, just as one has gauge-invariant
charged states in QED ({\it e.g.}, the state of one electron alone in the
world with its long-range Coulomb tail). 
Indeed, with unbroken global gauge invariance, QED and QCD have in common
that any gauge invariant localized state must be
chargeless/colorless \cite{Nak90}. The
question is the extension to non-local states as approximated by local ones.
In QED this leads to the so-called charge superselection sectors \cite{Haa96},
and non-local physical states with charge arise.    
If in QCD, with  unbroken global gauge symmetry {\em and}  mass gap, {\em
every} gauge-invariant state can be approximated by gauge-invariant localized
ones (which are colorless), one concludes that {\em every} gauge-invariant
(BRS-singlet) state must also be a color singlet.  

\subsection{Infrared dominance of ghosts in Landau gauge}

The (2nd condition in the) Kugo-Ojima confinement criterion,
$u = -\ID$ leading to well-defined charges $N^a$, can in Landau gauge be
shown by standard arguments employing Dyson-Schwinger
equations (DSEs) and Slavnov-Taylor identities (STIs) to be 
equivalent to an infrared enhanced ghost propagator \cite{Kug95}.
In momentum space the non-perturbative ghost propagator of Landau gauge QCD  
is related to the form factor occurring in the correlations of
Eq.~(\ref{Corru}) as follows,
\begin{equation}
    D_G(p) = \frac{-1}{p^2}      \, \frac{1}{ 1 + u(p^2) } \, , \;\;
                 \mbox{with}  \; \;   
                 u^{ab}(p^2)  = \delta^{ab}  u(p^2) \, . \label{DGdef}
\end{equation}
The Kugo-Ojima criterion, $u(0) = -1 $, thus entails that the Landau gauge
ghost propagator should be more singular than a massless particle pole in the
infrared. Indeed, there is quite compelling evidence for this exact
infrared enhancement of ghosts in Landau gauge \cite{Sme00}. 
For lattice calculations of the Landau gauge ghost propagator, see
Refs.~\cite{Cuc01,Cuc97,Sum96}. The Kugo-Ojima confinement criterion was also
tested on the lattice directly \cite{Nak99}.

Lattice verifications of the positivity violations for transverse gluon states
by now have a long history \cite{Man87,Mar95,Nak95,Ais96,Ais97,Man99}. 
Numerical extractions of their indefinite spectral density from lattice data 
are reported in \cite{Lan01}. As mentioned, however, this follows from color
antiscreening and superconvergence in QCD already in perturbation theory
\cite{Oeh80,Nis94}, and it is independent of confinement.  

Its remaining dynamical aspect resides in the cluster decomposition property
of local quantum  field theory in this formulation \cite{Haa96,Nak90}. 
Within the indefinite inner product structure of covariant QCD
it can be avoided for colored clusters, only {\em without mass gap} 
in the full indefinite space $\mathcal{V}$.   
In fact, if the cluster decomposition property holds for a gauge-invariant
product of two (almost local) fields, it can be shown that both fields are 
gauge-invariant (BRS-closed) themselves. 
With mass gap in the physical world,
this then eliminates the possibility of scattering a physical asymptotic 
state into a color singlet consisting of widely separated colored 
clusters (the  ``behind-the-moon'' problem) \cite{Nak90}. 

The necessity for the absence of the massless particle pole in $\partial_\nu
F^a_{\mu\nu} $ in the Kugo-Ojima criterion shows that the (unphysical)
massless correlations to avoid the cluster decomposition property are {\em
not} the transverse gluon correlations. An infrared suppressed propagator for
the transverse gluons in Landau gauge confirms this condition. This holds
equally well for the infrared vanishing propagator obtained from
DSEs \cite{Sti95,Sme97,Sme98}, 
and conjectured in the studies of the implications of the
Gribov horizon \cite{Gri78,Zwa92}, 
as for the infrared suppressed but possibly finite ones extracted from
improved lattice actions for quite large volumes \cite{Bon01,Bon00,Lei98}. 

An infrared finite gluon propagator with qualitative similarities 
in the transverse components appears to result also in simulations
using the Laplacian gauge \cite{Ale01}.   
Related to the Landau gauge, this gauge fixing was proposed as an
alternative for lattice studies in order to avoid Gribov copies
\cite{Vin92}. For a perturbative formulation see Ref.~\cite{vBa94}. Due to 
intrinsic non-localities, its renormalizability could not be
demonstrated so far. Deviations from the Landau gauge
condition were observed already at $O(g^2)$ in the bare coupling in
Ref.~\cite{Man01}. Moreover, the gluon propagator 
was seen to develop a large longitudinal component
in the non-perturbative regime \cite{Ale01}. In fact, compared to the
transverse correlations, it seems to provide the dominant component in the
infrared. And it might in the end play a role analogous to that of the
infrared enhanced ghost correlations in Landau gauge.   
However, the precise relation with Landau gauge still seems somewhat
unclear. It is certainly encouraging nevertheless to first of all verify 
that no massless states contribute to the transverse gluon correlations of
the Laplacian gauge either.

\subsection{Non-perturbative Landau gauge} 

A problem mentioned repeatedly already, which is 
left in the dark in the description of confinement within the covariant
operator formulation presented so far, is the
possible influence of Gribov copies~\cite{Gri78}.

Recently, renewed interest in stochastic quantisation arose, because  
it provides ways of gauge fixing in presence of Gribov copies, 
at least in principle \cite{Bau00,Bau01}. 
The relation to Dyson-Schwinger equations 
is provided by a time-independent version of the diffusion 
equation in this approach in which gauge-fixing is replaced by a globally
restoring drift-force tangent to gauge orbits in order to prevent the
probability distribution from drifting off along gauge orbit directions. 

In particular, in the limit of the Landau gauge, it is the conservative part
of this drift-force, the derivative w.r.t. transverse gluon-field components
of the Faddeev-Popov action, which leads to the standard Dyson-Schwinger
equations as clarified by Zwanziger \cite{Zwa01}.
He furthermore points out that  
these equations are formally unchanged, if Gribov's original suggestion 
to restrict the Faddeev-Popov measure to what has become
known as the interior of the first Gribov horizon is implemented.
This is simply because the Faddeev-Popov measure vanishes there,
and thus no boundary terms are introduced in the derivation of 
Dyson-Schwinger equations by this additional restriction. Phrased otherwise, 
it still provides a measure such that the expectation values of total
derivatives w.r.t. the fields vanish, which is all we need to formally
derive the same Dyson-Schwinger equations as those without restriction. 

In the stochastic formulation this restriction arises naturally because 
the probability distribution gets concentrated on the (first)
Gribov region as the Landau gauge is approached. 
Therefore, there should be no problem of principle with
the existence of Gribov copies in the standard DSEs.
However, the distribution of the probability measure
among the gauge orbits might be affected by neglecting (the non-conservative)
part of the drift force. Ways to overcome this approximation are currently
being investigated. Moreover, providing for a correct counting of gauge
copies inside the Gribov region, the full stochastic equation will 
allow comparison with results from Monte-Carlo simulations using lattice
implementations of the Landau gauge in a much more direct and reliable way. 
In particular, this should be the case for the lattice analog of the
stochastic gauge fixing used in simulations such as those of
Refs.~\cite{Nak95,Ais96,Ais97}.  

Here, we restrict to the standard Landau gauge DSEs which are best justified 
non-perturbatively from the stochastic approach to be valid modulo the
aforementioned approximation. For their solutions, on the other hand, 
restricting the support of the Faddeev-Popov measure to the interior of the
Gribov region has the effect of additional boundary conditions to 
select certain solutions from the set of all possible ones which might
contain others as well.
Consider two invariant functions $Z(k^2)$ and $G(k^2)$ to 
parameterize the Landau gauge structure,
\begin{eqnarray} 
        D_{\mu\nu}(k) = \frac{Z(k^2)}{k^2}  \left( \delta_{\mu\nu} -
        \frac{k_\mu k_\nu}{k^2} \right)  \, ,\;
        D_G(k)  = - \frac{G(k^2)}{k^2}          \, ,
\label{ZGdef}
\end{eqnarray}
in Euclidean momentum space of gluon and ghost propagator, respectively. 
When obtained as DSE solutions, suitable boundary conditions have to be
satisfied by these functions, in addition. 
The following infrared bounds were derived by Zwanziger
for each of the two as properties of the propagators from the restricted
measure: 

The observation that the ``volume'' of configuration space in the
infinite-dimensional (thermodynamic) limit is contained in its surface 
lead to the so-called {\em horizon condition} which entails that the ghost
propagator must be more singular than a massless particle pole in the
infrared \cite{Zwa01,Sch94,Zwa93},
\begin{equation}
       \lim_{k^2\to 0} G^{-1}(k^2) = 0 \; . \label{horizon}
\end{equation}
This condition is equivalent to the Kugo-Ojima criterion, $u = -\ID$ 
for well-defined color charges in Landau gauge, {\it c.f.}, Eqs.~(\ref{KO1})
and (\ref{DGdef}) with $G(k^2) = 1/(1+u(k^2))$.  

From the proximity of the Gribov horizon in infrared directions Zwanziger 
furthermore concluded \cite{Zwa92} that 
\begin{equation}
       \lim_{k^2\to 0} Z(k^2)/k^2   = 0 \; . \label{IRvan_glp}
\end{equation}
This removes the massless transverse gluon states of perturbation theory as
also required by the Kugo-Ojima criterion. The infrared vanishing of the
gluon propagator is a stronger requirement than this, however. 
It currently remains an open question why this has not been seen in
Monte-Carlo simulations as yet.  An infrared suppression of the gluon
propagator itself, rather than $Z(k^2)$, was observed for Landau gauge 
in \cite{Nak98} and, more considerably, 
at large volumes in $SU(2)$ in the 3-dimenional case
\cite{Cuc99,Cuc01a,Cuc01b}, as well as in Coulomb gauge \cite{Cuc00}. The
3-dimensional results are interesting in that the large distance gluon
propagator measured there seems incompatible with a massive behavior at low
momenta (that was noted also in \cite{Ais97}).  
At very large volumes, it even becomes negative \cite{Cuc01a,Cuc01b}. 
This is the same qualitative behavior as obtained for the one-dimensional
Fourier transform of the DSE results of \cite{Sme98,Sme97} at small values
for the remaining momentum components, {\it c.f.}, Fig.~4 of \cite{Sme00}
vs.~Figs.~2 of \cite{Cuc01a} or 6 of \cite{Cuc01b}. 
Qualitatively, the different dimensionality should not matter much here. 
On the other hand, the  extrapolation of the zero momentum
propagator in \cite{Cuc01b} leads to a finite result which, 
however, still decreases (slowly) with the volume. 
This suggests the physical volumes may still be too small yet and that
further study of the volume dependence of the zero momentum gluon propagator
might be necessary \footnote{The problem might be due to zero-momentum modes
yielding a volume dependent finite contribution which only dissapears in the
infinite volume limit, see also \protect\cite{Dam98}.}.

\subsection{Infrared exponents in previous studies}

Within the standard BRS or Faddeev-Popov formulation, the
functions in (\ref{ZGdef}) have been studied from
Dyson-\-Schwin\-ger equations (DSEs) for the 
propagators in various truncation schemes of increasing 
levels of sophistication \cite{Alk01,Sch01}.  
Typically, the known structures in the 3-point vertex functions, most
importantly from their Slavnov-Taylor identities and exchange symmtries, 
have thereby been employed to establish closed systems of non-linear integral
equations that are complete on the level of the gluon, ghost and quark
propagators in Landau gauge. This is possible with systematically neglecting
contributions from explicit  4-point vertices to the propagator 
DSEs as well as non-trivial 4-point scattering kernels 
in the constructions of the 3-point vertices \cite{Alk01,Sme98}.  
Employing a one-dimensional approximation, numerical 
solutions were then obtained in Refs.~\cite{Sme98,Sme97}.  

Asymptotic expansions techniques were developed to analytically
study the behavior of the solutions in the infrared.
The leading infrared behavior was thereby determined 
by one unique exponent $\kappa \approx 0.92 $,
\begin{equation} 
   Z(k^2) \, \stackrel{k^2\to 0}{\sim}   
      \,  \left(\frac{k^2}{\sigma}\right)^{2\kappa}  \mbox{and} \;\;
          G(k^2) \, \stackrel{k^2\to 0}{\sim}    
         \, \left(\frac{\sigma}{k^2}\right)^{\kappa} \; , \label{IRB}
\end{equation}
with a renormalization group invariant $\sigma $, see
Sec.~\ref{iarg}. 
The general bounds $0<\kappa <1$ were established in Ref. \cite{Sme98}
based on the additional requirement that $Z$ and $G$ have no zeros or poles
along the positive real axis, {\it i.e.}, in the Euclidean domain. Below, we
will verify the positivity for the leading infrared behavior of both these
functions in the same range, independent of the one-dimensional
approximation, and  based on some few and quite generic properties of the
ghost-gluon vertex alone.

The infrared behavior in Eqs.~(\ref{IRB})
was later confirmed qualitatively by studies of further truncated
DSEs. In Ref. \cite{Atk97}, a tree-level ghost gluon vertex was used 
in combination with a one-dimensional approximation which lead to a value of
$\kappa \approx 0.77$ for the infrared exponent of ghost and gluon
propagation in Landau gauge. Then, in the first infrared asymptotic study of
the ghost-gluon system without one-dimensional approximation, the value of
$\kappa = 1$ was obtained in Ref. \cite{Atk98}. There is, however, an issue
about infrared transversality of the gluon propagator, as we will explain
below, which was not addressed in this study. As a result, the correct value
for the tree-level vertex is the same as that derived herein 
for any ghost-gluon vertex with regular infrared limit, $\kappa \approx
0.595$, which was first reported for the tree-level vertex 
independently in Refs.~\cite{Ler01,Zwa01}.
As we furthermore find in our present study, inconsistency arises for 
$\kappa \to 1$ (from below), and this limit, the upper bound on $\kappa $,   
is therefore excluded.   

With $1/2 < \kappa $, all these values of the infrared exponent share,
however, the same qualitative infrared behavior. The gluon propagator
vanishes while the ghost propagator is infrared enhanced. Then, the
Kugo-Ojima criterion and the boundary conditions (\ref{horizon}),
(\ref{IRvan_glp}) are both satisfied. The horizon condition seems 
understandable, because the restriction to the Gribov region leads to a
positive measure which was implicitly also assumed by requiring solutions
without nodes in \cite{Sme98}. Here, depending on the infrared behavior of
the ghost-gluon vertex, we will find that this requirement could in principle
be maintained also for $\kappa < 1/2 $.  Taken by itself, it only leads to
$0<\kappa <1$, and thus to the horizon condition (\ref{horizon}).
Eventually, with decreasing $\kappa $ for values smaller than $1/2$,  
infrared singularities in ghost exchanges become too strong for a local field
theory description. Around $\kappa = 1/2$, however,  this argument is just
not strong enough, and we can not turn it into an independent additional 
argument in favor of (\ref{IRvan_glp}) for an infrared vanishing gluon
propagator.

\section{Dyson-Schwinger equations}

\begin{figure}[t]
\vspace{.2cm}
\epsfig{file=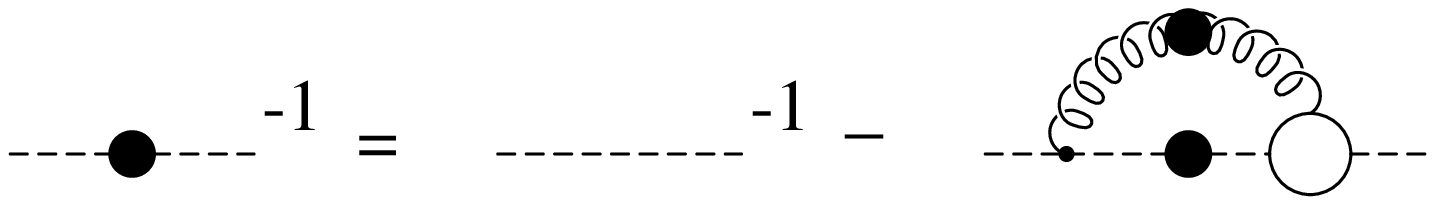,width=.81\linewidth}\hfill
\vspace{.2cm}
  \epsfig{file=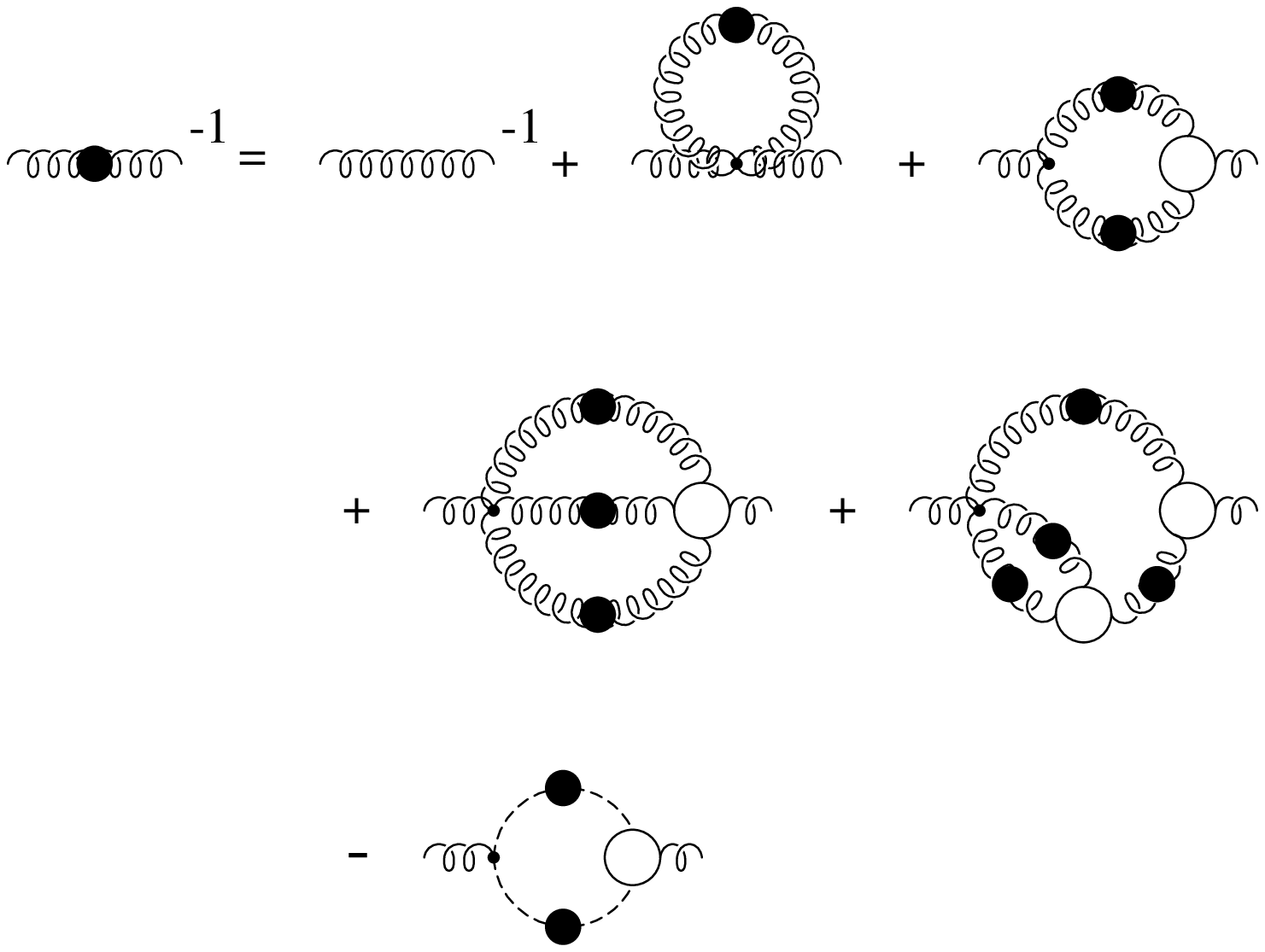,width=\linewidth}
\caption{Dyson-Schwinger equations for the ghost (top) and the gluon (bottom)
 propagator, diagrammatically.}  
\label{DSEfigs}
\end{figure} 

The Dyson-Schwinger equations for the propagators of ghosts and gluons 
in the pure gauge theory without quarks are schematically represented by
the diagrams shown in Fig.~\ref{DSEfigs}. With infrared dominance of ghosts,
the ghost loop represented by the diagram in the last line of
Fig.~\ref{DSEfigs} will provide the dominant contribution to the
inverse gluon propagator on the l.h.s. in the infrared.    
In our infrared analysis we will concentrate on this contribution to the
(renormalized) gluon DSE which reads in Euclidean momentum space 
with the notations of \cite{Alk01} (color indices suppressed),
\begin{eqnarray} 
  D^{-1}_{\mu\nu}(k)
    &=& Z_3 \, {D^0}^{-1}_{\mu\nu}(k) \,
 -\,   
   g^2 N_c \, \widetilde Z_1  \, \times
\label{glDSE}\\
 &&\hskip -1cm\int \frac{d^4q}{(2\pi)^4} \; G^0_\mu(q,p) \,
D_G(p)\, D_G(q)\, G_\nu(p,q)  \, + \cdots \nonumber
\end{eqnarray}
where $p = k + q$, and the contributions from the four remaining gluon
loop-diagrams of Fig.~\ref{DSEfigs} were not given explicitly.
$D^{0}$ is the tree-level propagator, $D_G$ is the ghost
propagator, and $G_\nu$ is the fully dressed ghost-gluon vertex function 
with its tree-level counter part denoted by $G^0_\mu $.
In the standard linear-covariant gauge 
the latter is given by the antighost-momentum, 
$G^0_\mu (q,p) = i q_\mu $. 
The DSE for the ghost propagator, without truncations at this point, 
formally reads, 
\begin{eqnarray}
  D_G^{-1}(k)  &=&  -\widetilde{Z}_3 \, k^2 \,   \label{ghDSE}\\
&& \hskip -1cm +\, g^2 N_c\,  \widetilde Z_1
  \int \frac{d^4q}{(2\pi)^4} \; ik_\mu \,
                     D_G(q) \, G_\nu (q,k) \, D_{\mu\nu}(k-q)   \; .
\nonumber
\end{eqnarray}
The renormalized propagators, $D_G$ and $D_{\mu\nu}$ and the
renormalized coupling $g$ are defined from the 
respective bare quantities by introducing multiplicative renormalization
constants, 
\begin{equation}
  \widetilde{Z}_3 D_G = D^{\hbox{\tiny bare}}_G \; , \quad
  Z_3 D_{\mu\nu} = D^{\hbox{\tiny bare}}_{\mu\nu} \; , \quad
  Z_g g = g_{\hbox{\tiny bare}} \; .
  \label{Zds}
\end{equation}
Furthermore, $\widetilde{Z}_1 = Z_g Z_3^{1/2} \widetilde{Z}_3$ is the
ghost-gluon vertex renormalization constant. Before we discuss the 
properties of the ghost-gluon vertex, essentially the only unknown in 
(\ref{glDSE}) and (\ref{ghDSE}), we note the following:   

If we are allowed to assume that the leading contribution to the 
inverse gluon propagator in (\ref{glDSE}) is 
completely determined by the ghost loop, this contribution must 
be transverse in Landau gauge. In other words, writing 
\begin{equation}
  D^{-1}_{\mu\nu}(k) =   k^2 \delta_{\mu\nu}  Z_P^{-1}(k^2) 
  \, - \, {k_\mu k_\nu}  Z_R^{-1}(k^2)     \; , 
\end{equation} 
one should then have $Z_P(k^2) = Z_R(k^2) \equiv Z(k^2)$. Here, in particular, 
the leading infrared behavior as extracted from the ghost loop alone
should not depend on whether we study $Z_P $ or $Z_R $. With all other
contributions subleading, deviations from the transversality of the
"vacuum polarization", $Z_P = Z_R$ in Landau gauge, should also be
subleading. We will assess this by studying in parallel (we sometimes  
use $D$ dimensions, normally $D\!=\!4$ here), 
\begin{eqnarray}
   Z_P^{-1}(k^2) &=&   \frac{1}{(D\!-\!1)k^2} \;  D^{-1}_{\mu\nu}(k)
   \mathcal{P}_{\mu\nu}(k)  \; , \label{eq:ZP} \\
   Z_R^{-1}(k^2) &=&   \frac{1}{(D\!-\!1)k^2} \;  D^{-1}_{\mu\nu}(k)
   \mathcal{R}_{\mu\nu}(k)  \; ,  \label{eq:ZR} \\
  && \hskip -2.3cm  \mbox{with} \;\;  \mathcal{P}_{\mu\nu}(k)  =
   \delta_{\mu\nu} -  \frac{k_\mu k_\nu}{k^2} \; ,  
 \; \; \mathcal{R}_{\mu\nu}(k)  =
   \delta_{\mu\nu} -  D \frac{k_\mu k_\nu}{k^2} \; ,\nonumber
\end{eqnarray}  
respectively. Beyond the leading behavior as dominated by the
infrared enhanced propagators within the ghost loop, there will in general be
complicated cancellations between longitudinal contributions from various
sources to ensure transversality of the gluons in Landau gauge. These 
sources can be due to the terms neglected, to truncations of vertices or 
to the regularization scheme employed. The tadpole for example contributes
only to $Z^{-1}_P$, and so do the quadratic divergences with cutoff
regularization. Beyond the leading order one therefore usually employed the
$\mathcal{R}$-tensor in the contraction of the gluon DSE in most
previous studies since that of Ref.~\cite{Bro89}.  

The tadpole contribution is a momentum-independent constant, so that it will
necessarily be subleading as compared to the infrared singular ghost-loop,
whenever that singularity is strong enough to lead to an infrared-vanishing
gluon propagator, or $Z(k^2) \sim (k^2)^{2\kappa}$ with $\kappa > 1/2$ 
for $ k^2\to 0$. 

The infrared analysis we present below is independent of the
regularization and both these reasons in favor of the $\mathcal{R}$-tensor
do therefore not apply to our present study. Nevertheless, even with ghost
dominance, exact transversality will in general only be
obtained by including all different structures possible 
in the ghost-gluon vertex that can contribute to the leading 
infrared behavior.

\subsection{The Ghost-Gluon Vertex in Landau gauge}
\label{ggvlg}

\begin{figure}[t]
  \epsfig{file=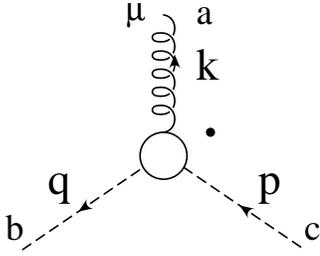,width=0.5\linewidth}
\caption{Conventions for the ghost-gluon vertex $G^{abc}_\mu(k,q,p)$.} 
\label{GluonGhostVertex}
\end{figure}

The ghost-gluon vertex is of particular importance in
the analysis of the infrared behavior of the gluon and ghost propagators.
We adopt the conventions of Ref.~\cite{Alk01}. 
The arguments of the ghost-gluon vertex denote in the following order 
the two outgoing momenta for gluon and ghost, and one incoming ghost
momentum, {\it c.f.}, Fig.~\ref{GluonGhostVertex}, 
\begin{eqnarray}
     G^{abc}_\mu(k,q,p) &=&  (2\pi)^4 \delta^4(k+q-p) G^{abc}_\mu(q,p) 
    , \,  \label{ghglvdef} \\
     G^{abc}_\mu(q,p) &=&   g f^{abc} G_\mu(q,p) \; .
\end{eqnarray}
Color structures other than the perturbative one assumed 
here were assessed for the pure Landau gauge theory 
on the lattice in Ref.~\cite{Bou98}. In this study, there 
was no evidence seen for any significant contribution due to 
such structures which we will not consider henceforth.   

We parametrise the general structure of $G_\mu(q,p)$ which consists of two 
independent terms by the following form,
\begin{equation}
\label{eq:GhGlV}
       G_\mu(q,p) = i q_\mu A(k^2;p^2,q^2) \, + \, ik_\mu B(k^2;p^2,q^2) \; .  
\end{equation}
One might expect the second structure to be insignificant in Landau gauge,
since it is longitudinal in the gluon momentum $k$. This is not necessarily 
the case in Dyson-Schwinger equations, however, since the transversality   
of the vacuum polarization generally arises from cancellations of different
longitudinal contributions as we discussed above. 

For later reference, we recall two general properties 
of this vertex. The implications of these properties are 
explored below. They both refer to the ghost-gluon vertex 
in \underline{Landau gauge:}
\begin{description} 
\item[(N)~]  Non-renormalization, $\widetilde Z_1 = Z_g Z_3^{1/2} 
            \widetilde Z_3 = 1$ 
            \cite{Tay71}, which entails that the vertex 
            reduces to its tree-level form
            at all symmetric momentum points, 
            \[ G_\mu(q,p) \Big|_{k^2=q^2=p^2} \, = \, G^0_\mu(q,p)  \; , \]
            in a symmetric subtraction scheme. The
            gauge fields being purely transverse, however, 
            there is a certain freedom left in the definition of the
            tree-level vertex. {\it A priori}, any form with $\eta \in [0,1]$, 
            \[\hskip 1cm 
              G^0_\mu(q,p) \, = \, \eta\, iq_\mu + \hat\eta \, i p_\mu \; ,
            \;\;   \mbox{and} \;\;  \eta + \hat\eta  = 1 \, , \]
            may be used equally for the Landau gauge. Without further
            specification, for the functions in 
            (\ref{eq:GhGlV}) we first have, 
            \[ A(x;x,x) \, = \, 1 \; , \;\; \mbox{and} \;\; B(x;x,x)= \hat\eta
            \; . \]
\end{description}            
The condition on $A$ is $\eta$-independent. It expresses 
the essential aspect of non-renormalization and will be
referred to as {\bf (N1)}. The condition on $B$ depends on  
the ambiguity in defining the Landau gauge, as expected. 
We call this condition {\bf (N2)}. 
It reads $B(x;x,x)=0$ for the transverse limit of the 
linear-covariant gauge in standard Faddeev-Popov theory, 
as compared to $B(x;x,x)=1/2$ for the analogous limit of the 
ghost-antighost symmetric Curci-Ferrari gauge \cite{Cur76}, see also
Ref.~\cite{Alk01}. These are the two special choices of particular interest,
corresponding to $\eta=1$ and $\eta=\hat\eta=1/2$, respectively.
For renormalizability and perturbative aspects of 
the latter, and for the geometry of the general $\eta $-gauges, see
Refs.~\cite{Bau82}.   
\begin{description} 
\item[(S)~]  The Ghost-antighost conjugation as part of the full Landau gauge
            symmetry, a semidirect product of $SL(2,\RR )$ and double BRS 
            invariance, implies \cite{Alk01},
            \[  A(x;y,z) \, = A(x;z,y) \; .   \hskip 1cm \mbox{\bf (S1)}
            \hskip -2cm\]
            While this holds for all $\eta$, again, 
            the $B$-function is more ambigous. It can not have definite
            symmetry properties in general.  
            For the symmetrized Landau gauge, based on the symmetric 
            tree-level vertex with $\eta\!=\!\hat\eta\!=\!1/2$, the   
            interactions with purely transverse gluons will preserve 
            this exact symmetry of the Landau gauge, however.  
            In the symmetric formulation we therefore expect to 
            have an exactly ghost-antighost symmetric vertex also,
            \[        G_\mu(q,p) \,=\,     G_\mu(p,q) \; .\]       
            Decomposing $B = B_+ + B_- $ with $B_\pm(x;z,y) = \pm
            B_\pm(x;y,z)$, we then furthermore deduce,
            \[     2B_+(x;y,z) =   A(x;y,z) \; .  \hskip 1cm \mbox{\bf (S2)}
            \hskip -2cm \]
\end{description} 
In the fully ghost-antighost symmetric formulation we can thus
express the vertex (\ref{eq:GhGlV}) in terms of the 
functions with definite (anti)symmetry as follows,  
\begin{equation}
\label{eq:GhGlVs}
       G_\mu(q,p) = \frac{i q_\mu + ip_\mu}{2} 
 A(k^2;p^2,q^2) +  ik_\mu B_-(k^2;p^2,q^2) \, .  
\end{equation}
The $B_-$-structure is absent at tree-level and it vanishes at all symmetric
points. Therefore, a symmetric vertex in converse to the logic 
above also requires $\eta=\hat\eta=1/2$ for the tree-level vertex to be used
as the $G^0_\mu$ in a symmetric subtraction scheme according to {\bf (N)}.

At this point, it seems important to stress that, for the infrared exponent
of Landau gauge QCD, only the infrared behavior of $A$ is relevant.
The critical exponents are of course independent of $\eta$.
As long as we concentrate on $Z_P$ via~(\ref{eq:ZP}) in the gluon DSE
(\ref{ghDSE}), the gluon legs of all vertices are transversely
contracted. For all internal gluon lines this is automatically true by 
the transversality of the propagator as in the ghost DSE (\ref{ghDSE}) for
example, and for the external lines we just arranged it by hand.
Thus, the $\eta$-freedom in the tree-level vertex and the $B$-structure 
of the full vertex are both irrelevant, as they should. In an infrared
anaysis based on this manifestly transverse system we might as well
have standard Faddeev-Popov theory in mind with $\eta=1$, $\hat\eta=0$.
  
The only place where the $\eta$-dependence and the $B$-structure do enter 
is the $\mathcal{R}$-contracted gluon DSE. We therefore introduce the
generalized Landau gauge by the above modification of the tree-level vertex 
here as a purely technical tool to address the transversality issue, {\it
i.e.}, to compare $Z_P $ and  $Z_R$ as obtained 
from Eqs.~(\ref{eq:ZP}) and (\ref{eq:ZR}), respectively. 
In fact, in order to reconcile ghost dominance with transversality, the
result will be that for arbitrary values of $\eta$ one must essentially have  
in the infrared (indicated by the superscripts),    
\begin{eqnarray}
B_+^{ir}(k^2;p^2,q^2) &=& \frac{1}{2} \, A^{ir}(k^2;p^2,q^2)
\; , \label{S2ir}\\ 
B_-^{ir}(k^2;p^2,q^2) &=& - \, \frac{p^2 - q^2}{2k^2} \,  A^{ir}(k^2;p^2,q^2)
\; .  \label{Bmir}
\end{eqnarray}
Inserting this into (\ref{eq:GhGlV}), small rearrangements 
reveal that the full ghost-gluon vertex therefore has to be transverse in the
infrared itself,     
\begin{equation}
       G^{ir}_\mu(q,p) \, = \,  \frac{i q_\mu\,  pk \, - \,  ip_\mu \,
                  qk}{k^2}  \,  A^{ir}(k^2;p^2,q^2) \, .  
\label{irGhGlVt}
\end{equation}
This is in contrast to its perturbative limit where the $B$-structure is
suppressed. And it is now also independent of the choice of $\eta$. 
Again, however, the transverse vertex is necessarily symmetric $\propto
(iq_\mu+ip_\mu)$ at a symmetric point. In order to extend the subtraction
scheme of {\bf (N)} non-perturbatively into the infrared, and ensure  
transversality of the gluon propagator, we thus have to
resort to the symmetric choice $\eta=\hat\eta=1/2$ also (for which
(\ref{S2ir}) follows trivially with {\bf (S2)}).

%

\subsection{Truncated Slavnov-Taylor identity}
\label{subsectSTI}

In \cite{Sme98} a Slavnov-Taylor identity of the 
standard linear-covariant gauge was derived to constrain 
the ghost-gluon vertex. Since the BRS transformations 
need some adjustments for other choices such as the ghost-antighost 
symmetric gauges, as it stands this identity is valid
for the case $\eta=1$, $\hat\eta=0$ only. 
A generalization might be worthwhile pursuing. However, considering   
the transversality of the full vertex in the infrared, this will not  
provide much additional information to be used in our present study, 
as we discuss in this subsection. 

Neglecting  irreducible ghost-ghost scattering contributions to the 
Slavnov-Taylor identity of Ref.~\cite{Sme98}, and thus maintaining the
disconnected contributions to the ghost 4-point 
function only, 
a truncated Slavnov-Taylor identity is obtained which, in terms of
the two structures $A$, $B$ in the vertex and the ghost propagator, reads, 
\begin{eqnarray}
  G(x) \Big(\frac{z\!+\!y\!-\!x}{2} A(y;x,z) - y B(y;x,z) \Big) +  && 
            \label{tSTI} \\
 G(y) \Big(\frac{z\!+\!x\!-\!y}{2} A(x;y,z) - x B(x;y,z) \Big) \!\!&=&\!\!
 \frac{zG(x)G(y)}{G(z)} \,. \nonumber
\end{eqnarray}
Without the symmetry property {\bf (S1)},
a simple solution to Eq.~(\ref{tSTI}) is given by
\begin{equation} 
  \label{vertEHW}     
   A(x;y,z) =  \frac{G(x)}{G(z)}    \; , \; \;   B(x;y,z) \equiv 0 \; .
\end{equation}   
This exact form was used for the ghost-gluon vertex in the study of Wilsonian
flow equations for Yang-Mills theory in Ref.~\cite{Ell98}. 
Implementing {\bf (S1)} in addition, the most general solution to
(\ref{tSTI}) can be written in the form,
\begin{eqnarray}
\label{eq:genSTIsol}
      A(x;y,z) &=&  \frac{G(x)}{G(z)} + \frac{G(x)}{G(y)} - 1 + x
      \, f^T\!(x;y,z) \; , \nonumber \\
  B(x;y,z) &=&  \frac{G(x)}{G(y)} - 1 + \frac{x\!-\!y\!+\!z}{2} 
   f^T\!(x;y,z) \; .\quad \label{ABgensol}
 \end{eqnarray}
The undetermined function $f^T$ thereby parametrises an unknown 
transverse contribution to the ghost-gluon vertex, $ k G^T\!(q,p) = 0 $,
of the typical type generally remaining unconstrained by the 
Slavnov-Taylor identities,  
\begin{equation} 
  G_\mu^T\!(q,p) = \big(iq_\mu  \, kp - i p_\mu \, kq \big) \,
  f^T\!(k^2;p^2,q^2) \; , \;\; 
\end{equation} 
where $ k=p\!-\!q $ as before. We obtain, however, from (\ref{ABgensol})
with {\bf (N1)} and {\bf (S1)}, 
respectively,
\begin{equation}
   f^T\!(x;x,x) =  0 \; , \;\; \mbox{and} \;\;
   f^T\!(x;y,z) = f^T(x;z,y) \; .  \label{fTc}
\end{equation} 
Since this function is otherwise arbitrary, in particular, in the infrared, 
the use of solutions (\ref{ABgensol}) somehow seems less
appealing for our present study which is concerned about the most general 
bounds on the infrared exponent that can be derived on the basis of as few
and basic assumptions as currently possible.   

In the numerical solutions to the coupled system of truncated 
ghost-gluon DSEs presented in Refs.~\cite{Sme97,Sme98}, the form given in
(\ref{ABgensol}) was used for the ghost-gluon vertex with $f^T\equiv 0$.  
This solution to the truncated STI (\ref{tSTI})  (for
$\eta=1$) still satisfies {\bf (N1)}, {\bf (N2)} and {\bf (S1)}. 
Because it is not purely transverse in the infrared,  
it should be used in combination with the transversely projected DSE for
$Z_P$ from (\ref{eq:ZP}), however, such that only the form of $A(x;y,z)$ in
(\ref{ABgensol}) matters.   
This causes ultraviolet problems in the numerical studies, see below. 
If the infrared transversality of the vertex can be maintained on the other
hand, by adding suitable transverse terms to a symmetric STI construction 
to satisfy (\ref{irGhGlVt}) above, for example, the $\mathcal{R}$-tensor may
be used to contract the gluon DSE via (\ref{eq:ZR}) by which these ultraviolet
problems are avoided without doing harm to the infrared structure of the
equations.  

We believe that this will be the way to proceed with the numerical studies of
full solutions to truncated DSEs in future. In particular, this suggests 
to further develop the ghost-antighost symmetric formulation.

\subsection{Ultraviolet subtractions and infrared behavior}

With the parametrisation of the vertex in (\ref{eq:GhGlV}) we now obtain
for the ghost-loop contribution to the gluon DSE (\ref{glDSE}) 
the two alternative expressions from the contractions according to
Eqs. (\ref{eq:ZP}) and (\ref{eq:ZR}), respectively,    
\begin{eqnarray}
\frac{1}{Z_P(k^2)} &=&  Z_3 + \frac{g^2N_c}{3}\int\frac{d^4q}{(2\pi)^4} \,
                \frac{G(p^2) G(q^2)}{k^2 p^2 q^2} \times  \label{ZPDSE}\\ 
        && \hskip 2.5cm     q\mathcal{P}(k)q \,  A(k^2;q^2,p^2) \, + \cdots 
                            \; , \nonumber  \\[4pt]
\frac{1}{Z_R(k^2)} &=&    Z_3 + \frac{g^2N_c}{3}\int\frac{d^4q}{(2\pi)^4} \,
                \frac{G(p^2) G(q^2)}{k^2 p^2 q^2} \times \label{ZRDSE} \\ 
        && \hskip -1.2cm \Big\{ \eta \big( q\mathcal{R}(k)p  \, A(k^2;q^2,p^2) 
      - \,  q\mathcal{R}(k)k  \, B(k^2;q^2,p^2) \big) + \nonumber\\
        && \hskip -1.8cm \hat\eta \big( p\mathcal{R}(k)p  \, A(k^2;q^2,p^2) 
      - \,   p\mathcal{R}(k)k  \, B(k^2;q^2,p^2)\big)\Big\} \, + \cdots 
                    \; , \nonumber
\end{eqnarray}
where again $p\!=\!k\!+\!q$. One can see explicitly here that knowledge of
both invariant functions is necessary for an infrared analysis based on 
the $\mathcal{R}$-tensor, {\it c.f.}, Eq.~(\ref{ZRDSE}), while only the
$A$-structure enters in Eq.~(\ref{ZPDSE}) obtained with the transverse
projector $\mathcal{P}$. We furthermore allowed for the generalized tree-level
vertex with $\eta+\hat\eta=1$ discussed in {\bf (N)} which does not affect
Eq.~(\ref{ZPDSE}). This both makes the equation for $Z_P^{-1}$ particularly
well suited for an infrared analysis, because then the invariant function 
$A(k^2;p^2,q^2)$, which parametrises the essential part of the vertex, is the
only  unknown remaining in the coupled system with the 
equally $\eta$-independent ghost DSE (\ref{ghDSE}),
\begin{eqnarray}
   \frac{1}{G(k^2)} &=&  \widetilde{Z}_3 -
   {g^2N_c}\int\frac{d^4q}{(2\pi)^4} \,  \frac{Z(p^2) G(q^2)}{k^2 p^2 q^2}
   \times  \hskip 1cm \label{GDSE} \\  
&&        \hskip 3cm     k\mathcal{P}(p)k  \,  A(p^2;k^2,q^2) \; .       
    \nonumber
\end{eqnarray}
We used $\widetilde{Z}_1 = 1$ for the Landau gauge in these equations. 
The ultraviolet divergences of the explicit loop-integrals are compensated by
the renormalization constants $Z_3$, $\widetilde{Z}_3$ which we require to 
be infrared finite. 
For the ultraviolet subtractions, which are of course $k$-independent, 
one then needs to make the following distinction: 

{\bf (i)} In DSEs for propagators of massless or infrared enhanced 
degrees of freedom we can perform the limit $k^2\to 0$. 
In the present case, we expect this for the ghosts, {\it i.e.}, the l.h.s.
of Eq.~(\ref{GDSE}) will approach a finite constant $ C_G \equiv \lim_{x\to 0}
G^{-1}(x) < \infty $ which can be zero, however. 
In such a case, the renormalization constant can easily be eliminated,
{\it e.g.}, here we then have,
\begin{eqnarray}
\label{ZGelim}
       \widetilde{Z}_3 &=& g^2 \mu^{4\!-\!D}  N_c \frac{D\!-\!1}{D} 
      \int\frac{d^Dq}{(2\pi)^D} \, \times \\
 && \hskip 1cm      \frac{1}{q^4} \, Z(q^2)G(q^2) A(q^2;0,q^2) \, + \, C_G\;. 
\nonumber
\end{eqnarray}
We adopted the conventions of dimensional regularization with $D\to 4^-$ 
here, but others may be employed as desired. 
If we assume regularity of $A$ in the origin, we can conclude
that $A(x;0,x) \to 1$ for $x\to 0$ from  {\bf (N1)}. Generally, if $A(x;0,x)$
does not vanish for $x\to 0$, then Eq.~(\ref{ZGelim}) tells us that
$Z(q^2)G(q^2) \to 0$  for $q^2 \to 0$ in the infrared. 
Otherwise, $\widetilde{Z}_3$ would be infrared divergent for $D\le 4$. 
On the other hand, $g^2Z(q^2)G^2(q^2) $ is more and more recognized
to be a good candidate for a non-perturbative definition of the running
coupling in Landau gauge, see Sec.~\ref{iarg}.  
However, this definition can only be
reasonable, if we are able to arrange matters such that it does not vanish in 
the infrared (in fact, the running coupling must be monotonic to
avoid a double valued $\beta$-function  with spurious zeros).

So that with  $Z(q^2)G^2(q^2) \not\to 0$ and $Z(q^2)G(q^2) \to 0$ we
must have infrared enhancement of ghosts. In particular, $C_G = 0$ in
Eq.~(\ref{ZGelim}) and no such constant is then possible in a
non-perturbative renormalization scheme. 

We can also reverse the logic here, and regard   
$C_G = 0$ as an additional boundary condition on a set of possible
solutions to the DSE (\ref{GDSE}).  This then implements Zwanziger's 
horizon condition (\ref{horizon}) to select the solution for the restricted 
Faddeev-Popov weight that vanishes outside the Gribov horizon \cite{Zwa01},
and that at the same time povides a positive definite $G(k^2) >0 $
\cite{Sme98}. Once this selection is made, however, 
it then follows that  $Z(q^2)G^2(q^2) \to const.$ for $q^2\to 0$
(in $D=4$ dimensions) which is a consequence of the non-renormalization of the
vertex {\bf (N1)}, as we will show in Sec.~\ref{IRghDSE}. So let us adopt
$C_G = 0$  and concentrate on the possible solutions with infrared
enhanced ghosts from now on. 

With (\ref{ZGelim}) in (\ref{GDSE}) we explicitly remove the 
ultraviolet divergence and obtain a manifestly finite equation  
to study the infrared behavior of $G$ in $D=4$ dimensions for a given 
form of $A$ in the infrared,
\begin{eqnarray}
   \frac{1}{G(k^2)} &=& {g^2N_c}\!\int\frac{d^4q}{(2\pi)^4} 
\, \Bigg\{ \frac{3}{4} \frac{Z(q^2)G(q^2)}{q^4} \,  A(q^2;0,q^2) 
    \nonumber\\
&& \hskip .2cm -   \,    k\mathcal{P}(p)k  \,   
 \frac{Z(p^2) G(q^2)}{k^2 p^2 q^2} \, A(p^2;k^2,q^2) \Bigg\} \; .       
  \hskip .5cm \label{GDSE4} 
\end{eqnarray}
While the ultraviolet subtraction is rather simple with $k^2 \to 0$ herein, 
without this subtraction, a naive infrared analysis will be aggravated by 
the ultraviolet divergences. Thus, the safe order of formal steps is 
to perform the ultraviolet subtraction before the infrared analysis  
in this case. The opposite order applies for the gluon DSE.

{\bf (ii)} In DSEs for propagators of massive degrees of freedom
or even infrared-vanishing correlations, the explicit ultraviolet
subtraction is subleading in the infrared, and it cannot simply be  
extracted from the limit $k^2\to 0$. This should certainly be 
the case for the transverse gluon correlations. The least we expect
as a necessary condition for confinement is the mass gap. The horizon
condition implies an even stronger infrared singularity, as mentioned in the
introduction. In either case we have $Z^{-1}(k^2) \to \infty$ for $k^2 \to 0$
for the l.h.s. in DSEs such as (\ref{ZPDSE}) or (\ref{ZRDSE}). The advantage
is that the coefficients of the divergent terms in an asymptotic infrared
expansion can be extracted without bothering about ultraviolet subtractions
in the first place, since ultraviolet divergences will only occur at the
subleading constant level in this expansion. To be specific, inside the ghost
loop, the infrared enhanced asymptotic forms of ghost propagators (together
the leading behavior of the ghost-gluon vertex) will converge rapidly in the
ultraviolet. Here the problem rather is to extract the necessary ultraviolet
subtraction without introducing by hand spurious infrared divergences. This
problem had already to be dealt with in Refs.~\cite{Sme97,Sme98}. What
generally needs to be done in such a case is to reverse the order of {\bf
(i)} above, and to isolate the infrared divergent contributions to the gluon
DSE on both sides prior to the ultraviolet subtraction, in order to define an
infrared finite renormalization constant $Z_3$.

Assume the coefficients in the 
infrared divergent terms of the asymptotic expansions for the ghost
propagator and the vertex are known and the corresponding asymptotic
forms denoted by $G^{ir}(q^2)$ and $A^{ir}(k^2;p^2,q^2)$. We can then, {\it
e.g.}, in Eq.~(\ref{ZPDSE}) subtract on both sides an ultraviolet finite
contribution of the form,
\begin{eqnarray}
\label{Zir}
    \frac{1}{Z_P^{ir}(k^2)} &\equiv&    
\frac{g^2N_c}{3}\! \int\frac{d^4q}{(2\pi)^4} \,
       \frac{q\mathcal{P}(k)q}{k^2 p^2 q^2}  \times  \\
&&   \hskip 1cm   G^{ir}(p^2) G^{ir}(q^2)   \, 
    A^{ir} (k^2;q^2,p^2) \; .
\nonumber\end{eqnarray}
which we will actually use to determine $Z_P^{ir}(k^2)$ below,
or an analogous expression for $Z_R^{ir}(k^2)$ from (\ref{ZRDSE}). 
In both cases, this infrared subtraction in the gluon DSE has to be performed
up to the necessary order such that 
\begin{equation}
\label{IRsubZ}
       \frac{1}{Z(k^2)} - \frac{1}{Z^{ir}(k^2)} \, \to \, C_A = const.\quad 
            \mbox{for} \quad k^2 \to 0 \;.  
\end{equation}
On the other hand, 
the $k$-independent constant contribution to the r.h.s. of the DSE
is the one that contains the (overall) logarithmic divergence which is
absorbed in the renormalization constant $Z_3$. 
To extract that from the infrared subtracted gluon DSE, 
analogous to 
Eq.~(\ref{ZGelim}) from the ghost DSE in {\bf (i)} above,
we first rewrite the gluon DSE~(\ref{ZPDSE}), adding a
zero via the definition in (\ref{Zir}),
\begin{eqnarray}
 Z_3&=& \frac{1}{Z_P(k^2)} - \frac{1}{Z_P^{ir}(k^2)} - 
    \frac{g^2\mu^{4\!-\!D}N_c}{D\!-\!1}\! \int\frac{d^Dq}{(2\pi)^D} \,
      \times  \nonumber\\
&&   \hskip -.4cm 
  \frac{q\mathcal{P}(k)q}{k^2 p^2 q^2} \, \Big( G(p^2) G(q^2)
    \,A(k^2;q^2,p^2)  \label{Z3mu} \\ 
&& \hskip 1cm  - \, G^{ir}(p^2) G^{ir}(q^2)   \, 
    A^{ir} (k^2;q^2,p^2) \Big)   \,+ \,\cdots \, .
 \nonumber
\end{eqnarray}
A further technical complication arises from the
additional power of $q^2/k^2$  which 
prevents us from taking the limit $k^2\to 0$ explicitly herein. 
This problem is due to contributions which superficially contain 
quadratic ultraviolet divergences.
Once the quadratic divergences have been
taken care of, whenever this is necessary in a given scheme,
we can then let $k^2 \to 0$ for the logarithmically divergent contribution
that defines the ultraviolet renormalization constant $Z_3$.  
For the $\mathcal{R}$-tensor, leading to $Z_R$ via Eq.~(\ref{eq:ZR}) which is
free from quadratically divergent contributions \cite{Bro89}, and for the 
generalized tree-level vertex $G_\mu = G^0_\mu $, as one special case to give
an example without such complication, one readily verifies that   
\begin{eqnarray}
 Z_3&=& - 
    \Big(\frac{2(D\!-2)}{D(D\!+2)} - \eta\hat\eta \Big) 
        \,   g^2\mu^{4\!-\!D}N_c \! \int\frac{d^Dq}{(2\pi)^D} \,
      \times  \nonumber\\
&&   \hskip .8cm 
  \frac{1}{q^4} \, \Big( G^2(q^2) - \big(G^{ir}(q^2)\big)^2  \Big)  \;
     + \,\cdots \, ,
 \label{Z3spec}
\end{eqnarray}
from $k^2\to 0$ in (\ref{ZRDSE}) with the analogue of (\ref{Zir}) for the
$\mathcal{R}$-tensor.  
It is infrared finite by construction, and it gives the correct perturbative
ghost-loop contribution to the gluon renormalization constant $Z_3$,
ultraviolet divergent for $D\to 4$. Contrary to the ghost DSE
renormalization in Eq.~(\ref{ZGelim}), an additive constant is possible in
(\ref{Z3spec}) and is included in the terms not given explicitly here.  
It can be used to adjust the constant $C_A$ in the gluon DSE 
which is subleading in the infrared, {\it c.f.}, (\ref{IRsubZ}).

To summarize,  it should be possible to impose 
renormalization conditions on the full propagators to 
equal the tree-level ones at an arbitrary (space-like) subtraction point
$k^2=\mu^2>0$.   
To do this, we have two independent conditions at our disposal 
which fix the physically insignificant overall factors in each of the 
two propagators. For massless tree-level propagators, 
we can\-not without loss of generality extend this subtraction scheme to
include $k^2=\mu^2 = 0$, however, since that forced the full propagators to  
have the same massless single-particle singularity. Here, the 
necessity of a mass gap in the transverse gluon correlations entails that 
$1/Z(k^2)$ is infrared divergent, and it is thus impossible to fix a
multiplicative factor by a subtraction at $\mu=0$ requiring that $1/Z(0) $ 
be unity (or any other finite value). We can, however, fix this factor 
by assigning the infrared subtracted $1/Z(k^2) - 1/Z^{ir}(k^2) $ 
at $k^2 =0$ a certain value $C_A$. 

That sets one of the two conditions available. Of course, the same argument 
applies to the ghost propagator at $\mu =0$. In this case, it is because both, 
the Kugo-Ojima criterion and the horizon condition, tell us that the full ghost
propagator should not have the singularity structure of the free-massless  
tree-level one. In particular, with $1/G(k^2) \to 0$ we cannot fix the
overall factor with subtracting $G^{-1}$ at zero.  As mentioned above, this
case is different in that a non-vanishing constant contribution to the ghost
DSE would be infrared dominant and cannot occur toghether with the infrared
enhanced ghost correlations. To fix the multiplicative factor in the ghost
propagator implicitly, we use the second of the two independent
renormalization conditions on the product of both propagators, 
\begin{equation}
G^2(\mu^2) Z(\mu^2) \, \stackrel{!}{=} \, 1 \; ,
\end{equation}
which can be used to define a non-perturbative running coupling in Landau
gauge as we discuss next.

\section{Renormalization independent infrared analysis}
\label{riia}

\subsection{Infrared expansion and renormalization group} 
\label{iarg}

Herein, we adopt the non-perturbative renormalization scheme introduced in
Refs.~\cite{Sme97,Sme98}. To  review this scheme briefly, recall that the 
formal solutions to the renormalization group equations for the gluon and the
ghost propagator, {\it e.g.}, for the latter this is Eq.~(\ref{RGG}) in
App.~\ref{corrRG}, can be written in the general forms,
\begin{eqnarray}
\hskip -.5cm  Z(k^2) \!\!&=& \!\! \exp\bigg\{\!\! -2 \int_g^{\bar g(t_k, g)} \!\!\!dl \,
  \frac{\gamma_A(l)}{\beta(l)} \bigg\} \, f_A(\bar g(t_k, g)) \; ,
  \label{RGsZ} \\ 
\hskip -.5cm   G(k^2) \!\! &=& \!\! \exp\bigg\{\!\!  -2 \int_g^{\bar g(t_k, g)}  \!\!\! dl \, 
  \frac{\gamma_G(l)}{\beta(l)} \bigg\} \, f_G(\bar g(t_k, g)) \; ,
  \label{RGsG}
\end{eqnarray}
respectively. Here, $t_k = (\ln k^2/\mu^2)/2$, and  $\bar g(t,g)$ is the
running coupling $\bar g(t,g)$, the solution of
$d/dt \, \bar g(t,g) = \beta(\bar g) $ with $\bar g(0,g) = g$ and with the
Callan-Symanzik $\beta$-function, perturbatively $\beta (g) = - \beta_0 g^3 +
{\cal O}(g^5)$. The exponential factors are the multiplicative constants for
finite renormalization group transformations ($\mu\to\mu'$), 
\begin{eqnarray}
\hskip -.5cm  
 \mathcal{Z}_3(\mu',\mu) \!\!&=& \!\! \exp\bigg\{\!\! -2 \int_g^{g'} \! dl
  \frac{\gamma_A(l)}{\beta(l)} \bigg\}  \; ,
  \label{ZZ} \\ 
\hskip -.5cm  
  \widetilde{\mathcal{Z}}_3(\mu',\mu) \!\! &=& \!\! \exp\bigg\{\!\!  -2
  \int_g^{g'} \! dl 
  \frac{\gamma_G(l)}{\beta(l)} \bigg\} \; ,
  \label{ZG}
\end{eqnarray}
with ${\mu'}^2 = k^2$ in (\ref{RGsZ}), (\ref{RGsG}),
where $\gamma_A(g)$ and $\gamma_G(g)$ are the anomalous dimensions of the
gluon and the ghost fields, respectively. 

The structure of (\ref{RGsZ}), (\ref{RGsG}) is summarized as follows:
The momentum dependence of the propagator functions $Z(k^2)$ and $G(k^2)$ 
is completely determined by the running coupling evaluated at $k^2$, which  
is renormalization group invariant, {\it i.e.}, $\mu$-independent, since 
$(d/d\ln\mu) \bar g(t_k,g) = ({\mu\partial}/{\partial \mu} + \beta(g)
{\partial}/{\partial g}) \bar g(t_k,g) =0$. We can therefore 
parametrise this momentum dependence by a function of the ratio of $k^2$ over
a renormalization group invariant, dynamically generated momentum scale
$\sigma \propto \Lambda^2_{\mbox{\tiny QCD}}$.  

The $\mu$-dependence of the propagators, on the other hand, is then 
given only by the $g\equiv g(\mu)$ of the lower bound in the exponential
renormalization factors. We can therefore always
separate these two dependences, 
that on $(g,\mu)$ versus that on $k^2/\sigma$, 
in (\ref{RGsZ}), (\ref{RGsG}) into multiplicative factors by conveniently
choosing a $g_0$ such that $\mu^2 =\sigma $ at $g=g_0$,
\begin{equation}
   \sigma = \mu^2 \exp\big\{
     -2\int^g_{g_0} \frac{dl}{\beta(l)} \big\} \; ,    \label{smug}
\end{equation} 
which, at the same time, defines $\sigma $ to be a renormalization group
invariant momentum scale as promised.  

With this factorization of the $(g,\mu)$-dependence, we now   
make the Ansatz that the propagator functions $Z$ and $G$ 
have asymptotic infrared expansions in terms of
$k^2/\sigma $ to some order $N$,
\begin{eqnarray}
\hskip -.4cm    Z(k^2)  &\to&   \exp\bigg\{\!\! -2 \int_g^{g_0} \!\! dl \,
  \frac{\gamma_A(l)}{\beta(l)} \bigg\} \,     \sum_n^N
    e_n    \left(\frac{k^2}{\sigma}\right)^{\epsilon_n}  \hskip -.2cm   , 
\label{asZ} \\
\hskip -.4cm   G(k^2)  &\to&   \exp\bigg\{\!\!  -2 \int_g^{g_0}  \!\! dl \, 
  \frac{\gamma_G(l)}{\beta(l)} \bigg\} \,  \sum_n^N
   d_n  \left(\frac{k^2}{\sigma}\right)^{\delta_n}   \hskip -.2cm   ,
\label{asG} 
\end{eqnarray} 
for $k^2/\sigma \to 0$. Here we use a notation similar to that 
introduced in \cite{Wat01}. We note, however, that our expansion 
involves the RG invariant scale $\sigma $ while the renormalization scale 
$\mu $ was used in \cite{Wat01}. The difference can be absorbed in a
redefinition of the coefficients $e_n$, $d_n$ as we explain in
App.~\ref{corrRG}.  Most importantly, this implies that our coefficients 
$e_n$, $d_n$ are also $(g,\mu)$-independent.

The non-renormalization {\bf (N)} of the ghost-gluon vertex in the infrared,
{\it c.f.}, $\widetilde{Z}_1 =Z_g Z_3^{1/2} \widetilde{Z}_3 = 1$ which was 
derived, in particular, for a symmetric subtraction scheme
$k^2=p^2=q^2=\mu^2$ with $\mu^2 \to 0$ \cite{Tay71}, now entails for the
renormalization factors in Eqs.~(\ref{ZZ}), (\ref{ZG}) that
\begin{equation} 
\mathcal{Z}_3^{1/2}(\mu',\mu) \widetilde{\mathcal{Z}}_3(\mu',\mu) = \frac{g'}{g} \equiv
\frac{\bar g(t,g)}{g}  \; , \;\; t=\ln\frac{\mu'}{\mu} \; ,  
\label{gpLg}
\end{equation} 
in Landau gauge, which is equivalent to 
\begin{equation} 
      2 \gamma_G(g) \, +\, \gamma_A(g)  \, = \, -\frac{1}{g} \, \beta(g)  \; .
  \label{andim}
\end{equation} 
This is, in fact, what allows to define a 
non-perturbative running coupling as introduced 
in Refs.~\cite{Sme97,Sme98} by,
\begin{equation} 
  g^2 Z({\mu'}^2) G^2({\mu'}^2) \, \stackrel{!}{=} \,  {g'}^2 = \bar g^2(\ln
(\mu'/\mu) , g) \; . \label{gbar}
\end{equation} 
It reduces to the unique perturbative definition for large $\mu$, $\mu'$,
is renormalization group invariant, dimensionless and thus as good as any
non-perturbative definition can be. The fact that no constant of 
proportionality is involved in (\ref{gbar}) implies a specific
renormalization condition. 
It corresponds to requiring the condition on the propagators,
\begin{equation}
Z(\mu^2) =  f_A(g) \;, \;\; G(\mu^2) =  f_G(g)
  \; \; \hbox{with} \;\; f_G^2 f_A = 1 \; ,
  \label{npsub}
\end{equation}
which is incomplete, of course, to fix both their values separately 
at an arbitrary $k^2 = \mu^2$. The perturbative limits are, however, 
$ f_{A,\, G} \to 1 \, , \; g \to 0 $, corresponding to the perturbative MOM,
\begin{equation}
  Z(\mu^2) = 1 \quad \mbox{and} \quad G(\mu^2) = 1
  \label{persub}
\end{equation}
for an asymptotically large subtraction point $k^2 = \mu^2$.

With this so defined running coupling, by Eq.~(\ref{gbar}), 
the existence of an infrared fixed point, 
$\bar g(t, g) \to g_c $ finite for $t\to -\infty$, 
then follows in Landau gauge 
to be in one-to-one correspondence with the scaling law for the leading
infrared exponents of gluon and ghost propagation 
in the form (for $D\!=\!4$),   
\begin{equation}
\label{sclaw}
  \epsilon_0 + 2\delta_0 = 0 \; .
\end{equation}
To make this explicit, consider the leading infrared behavior from
(\ref{asZ}), (\ref{asG}), with the exponential factors therein 
expressed by (\ref{ZZ}), (\ref{ZG}), for $k^2 \to 0$,
\begin{eqnarray}
\hskip -.4cm    Z(k^2)  &\to&    \mathcal{Z}_3(\sqrt{\sigma},\mu)\;  
    e_0    \left(\frac{k^2}{\sigma}\right)^{\epsilon_0}  \hskip -.2cm   , 
\label{aslZ} \\
\hskip -.4cm   G(k^2)  &\to&  \widetilde{\mathcal{Z}}_3(\sqrt{\sigma},\mu) \;
    d_0  \left(\frac{k^2}{\sigma}\right)^{\delta_0}   \hskip -.2cm   ,
\label{aslG} 
\end{eqnarray} 
which, from (\ref{gpLg}), entails that the infrared behavior of the running
coupling $\alpha(k^2) \equiv \bar{g}^2(t_k,g)/(4\pi)$ is given by,
\begin{equation}
\label{alpha_ir}
     \alpha(k^2) \equiv \frac{g^2}{4\pi} Z(k^2) G^2(k^2) \to 
     \frac{g_0^2 e_0 d_0^2}{4\pi} \, 
     \left(\frac{k^2}{\sigma}\right)^{\epsilon_0 + 2 \delta_0} .  
\end{equation}   
We therefore introduce $\alpha_c =  g_0^2 e_0 d_0^2/(4\pi)$ in
the following. It represents the infrared fixed point, 
$\alpha (k^2)  \to \alpha_c  $ for $k^2 \to 0$,  which occurs exactly if  
(\ref{sclaw}) holds. 

The infrared scaling behavior in Eq.~(\ref{sclaw}) was first observed in the
solutions to truncated DSEs in Refs.~\cite{Sme97,Sme98}. It
was recently derived from the ghost DSE in
Ref.~\cite{Wat01}. Therein, an additional assumption on the vertex
was used which is a bit  {\it ad hoc} and which is actually not necessary.
In Sec.~\ref{IRghDSE} below, we therefore present an alternative
derivation of the infrared behavior~(\ref{sclaw}), from Ref.~\cite{Ler01},
which is based on the non-renormalization, condition 
{\bf (N1)} for the vertex alone.

\subsection{Vertex Ansatz}
\label{subsecVA}
 
All we need
in our infrared analysis is an Ansatz for the 
invariant function $A$ which parametrises the relevant structure of
the ghost-gluon vertex in Landau gauge. 
Anticipating a conformal behavior in
the infrared also for the vertex, we first write, 
\begin{equation}
\label{AirAns}
    A^{ir}(k^2;p^2,q^2) = \left(\frac{k^2}{\sigma}\right)^n 
    \left(\frac{p^2}{\sigma}\right)^m   \left(\frac{q^2}{\sigma}\right)^l \;
    .    
\end{equation}   
The non-renormalization condition {\bf (N1)} for the Landau-gauge vertex
then leads to the constraint,
\begin{equation}
\label{expconst}
        l+m+n = 0  \; ,
\end{equation} 
which will be implemented in our analyses, throughout.
We will later also allow sums of terms of this form, in order to
explore versions of this Ansatz which are symmetrized w.r.t. the ghost legs. 

The scaling law for the infrared propagators in Eq.~(\ref{sclaw}) 
then follows for such a sum of terms  (\ref{AirAns}) via (\ref{expconst}) 
only from  {\bf (N1)}, as we shall show in the next subsection. 

If we require, in addition, that $A(k^2;p^2,q^2) $ 
remain finite when one of the ghost momenta vanishes, 
one of the terms of the form (\ref{AirAns}) must exist in the sum  
with either 
\begin{equation}
                m = 0 \; ,  \;\;  l = - n \; , \quad \mbox{or}\qquad 
                l =0  \;, \;\;  m = -n \; . 
      \label{addexpc}
\end{equation}
All other possible terms must then vanish and thus have $m>0$ or $l>0$,
respectively. And if the finite contribution to $A$ in
that limit was to be in itself symmetric w.r.t.~the two ghost
momenta, one would only be left with $A^{ir} \equiv 1$  as 
in the tree-level vertex, since then $l = m = 0$. In a ghost-antighost
symmetric sum of two terms on the other hand, we need one of each kind
together with $n<0$ to avoid infrared divergent ghost legs.

After the infrared scaling (\ref{sclaw}) and the general formulae for the
infrared contributions of terms  of the genuine form (\ref{AirAns}),
(\ref{expconst}) will be derived,  
we will assume relations as in (\ref{addexpc}), in addition.
The joint infrared exponent $\kappa$ for ghosts and
gluons then is a function
of a single  critical exponent $n$ which is left as an open parameter 
in their vertex.   
To exemplify its influence, we will explicitly calculate this function for
the following three Ans\"atze:
\begin{eqnarray}
\lefteqn {(i)}   
&&\qquad    A^{ir}(k^2;p^2,q^2) = \left(\frac{k^2}{q^2}\right)^n  
\label{verti} , \\
\lefteqn {(ii)}  
&&\qquad    A^{ir}(k^2;p^2,q^2) = \frac{1}{2} \left(
 \left(\frac{k^2}{q^2}\right)^n \!\! 
       +  \left(\frac{k^2}{p^2}\right)^n  \right) 
         \; ,\label{vertii} \\
\lefteqn {(iii)}  &&\qquad   
 A^{ir}(k^2;p^2,q^2) =  \left(\frac{k^2}{q^2}\right)^n \!\! 
 +  \left(\frac{k^2}{p^2}\right)^n\!\! - 1   \; .   \qquad \label{vertiii}
\end{eqnarray}
In the form $(i)$ of (\ref{verti}) the Ansatz does not bother about
ghost-antighost symmetry. For $n \!=\! \delta_0 \!\equiv\! - \kappa$, 
this form contains the infrared behavior of the non-symmetric solution
(\ref{vertEHW}) to the truncated STI (\ref{tSTI}) for the vertex of
Ref.~\cite{Ell98} as a special case. 
In both symmetrized versions $(ii)$ and $(iii)$ 
one furthermore has $n\le 0$, if infrared divergences associated with the
ghost legs are to be avoided in the ghost-gluon vertex function.
Version $(ii)$ in (\ref{vertii}) yields a finite and 
constant $A(q^2;0,q^2)= 1/2 $, while  $(iii)$ is an example for 
the possibility that $A(q^2;0,q^2) = 0$. 
Version $(iii)$ in Eq.~(\ref{vertiii}) for $n \!=\! \delta_0 \!\equiv\! -
\kappa$ includes the behavior of the symmetric solution (\ref{eq:genSTIsol})
to the STI (\ref{tSTI}) with $f^T\equiv 0$ as obtained in Ref.~\cite{Sme98}.
All three versions satisfy {\bf (N1)}, of course, and they all 
reduce to the tree-level vertex at $n=0$.

\subsection{Unique infrared exponent from ghost DSE}
\label{IRghDSE}

With the general form of our Ansatz (\ref{AirAns}) for the relevant part 
of the ghost-gluon vertex, we can now extract the leading contributions 
in the infrared on both sides of the ghost DSE (\ref{GDSE4}).
Here, with $Z^{ir}$ and $G^{ir}$ denoting 
the leading infrared behavior of the propagators as given in
(\ref{aslZ}) and (\ref{aslG}), respectively, we first note that 
\begin{equation} 
     Z^{ir}(p^2)G^{ir}(q^2) = \frac{g_0^2}{g^2}
\frac{ e_0\, d_0 }{\widetilde{\mathcal{Z}}_3(\sqrt{\sigma},\mu)}
     \left(\frac{p^2}{\sigma}\right)^{\epsilon_0}    
       \left(\frac{q^2}{\sigma}\right)^{\delta_0}  ,
\label{irprod} 
\end{equation}
where use has been made also of Eq.~(\ref{gpLg}) for $\mu'=\sqrt{\sigma}$,
$g'=g_0$. The l.h.s. of the ghost DSE (\ref{GDSE4}) approaches, for $k^2\to
0$,  
\begin{equation} 
    G^{-1}(k^2) \to  \widetilde{\mathcal{Z}}_3^{-1}(\sqrt{\sigma},\mu) \,
              d_0^{-1} \left(\frac{k^2}{\sigma}\right)^{-\delta_0} . 
\label{irlhsG}
\end{equation}
To obtain the leading behavior at small $k^2/\sigma$ of the r.h.s. in
(\ref{GDSE4}), we replace the undetermined functions in the integrand 
of (\ref{GDSE4}) by the from given in (\ref{irprod}) and the Ansatz
(\ref{AirAns}) for the leading infrared behavior of the vertex. 
The difference in the integrand between the full
functions, $Z$, $G$ and $A$, and their asymptotic infrared forms, $Z^{ir}$,
$G^{ir}$ and $A^{ir}$, is subleading and it produces, upon integration, terms
that are also subleading in an expansion of the r.h.s.~of the DSE. 
This procedure is not restricted to the leading infrared behavior. 
It can straightforwardly be generalized for an infrared expansion to a given
order, as long as all integrals in this expansion remain finite. 
This is true for the leading infrared forms as discussed above. 
When these are inserted, the integral in (\ref{GDSE4}) converges in $D=4$
dimensions. The role of the first term, however, just is to guarantee
convergence of all necessary integrals at exactly $D=4$. It is certainly 
necessary for the subtraction of the ultraviolet divergences 
when the full functions with their logarithmic momentum dependences are
inserted. Here we only need this term, if we insist on a calculation in 
$D=4$ involving convergent integrals in every step. We can, however,
obtain the same result by analytic continuation of integrals performed in
$D$ dimensions. Though this is of course precisely along the rules of
dimensional {regularization}, this notion might be misleading here, since
nothing is there to be regularized in the first place. It is adopted here 
for convenience only. Upon insertion of the leading infrared forms
(\ref{irprod}) and (\ref{AirAns}),
the first term in (\ref{GDSE4}), in $D$ dimensions, then yields a
contribution to the ghost DSE proportional to 
\begin{equation} 
         \left(\frac{k^2}{\sigma}\right)^{m} \int \frac{d^Dq}{(2\pi)^D} 
      \left(\frac{q^2}{\sigma}\right)^{\epsilon_0+\delta_0+l+n-2}
\end{equation} 
which vanishes for any $D$ in the analytic definition. 
      
Therefore, with (\ref{irprod}) and (\ref{AirAns}) in the integral for
relevant part of the r.h.s. in $D$ dimensions, and with (\ref{irlhsG}) for
the l.h.s., the leading contributions to both sides of the ghost DSE
(\ref{GDSE4}) for $k^2\to 0$ are readily extracted to yield,
\begin{eqnarray}
  \left(\frac{k^2}{\sigma}\right)^{\!\!-\delta_0}\hskip -.3cm &=&  
- N_c \, 4\pi\alpha_c \, 
\left(\frac{k^2}{\sigma}\right)^{\!\!\epsilon_0+\delta_0+\frac{D}{2}-2+l+m+n}  
 \hskip -.2cm \times \qquad\quad
\label{GDSEirD} \\
&&  \hskip -1.3cm    \int\!\frac{d^Dq}{(2\pi)^D}  
 \left(\frac{1}{q^2}\right)^{\!\!\frac{D}{2}}  \frac{k\mathcal{P}(p)k}{k^2}
        \left(\frac{q^2}{k^2}\right)^{\!\!\frac{D}{2}-1+\delta_0+l}  
     \left(\frac{k^2}{p^2}\right)^{\!\!1-\epsilon_0-n} \hskip -.3cm .
 \nonumber 
\end{eqnarray}
Here, $p=k\pm q$ again, and we have $ 4\pi\alpha_c =  g_0^2 e_0 d_0^2$ as
introduced in (\ref{alpha_ir}) of Sec.~\ref{iarg}. And just as for the usual 
replacement $g \to g \mu^{2-D/2}$, the dimension of the coupling for 
$D\!\not\!=4$ has been taken care of in (\ref{GDSEirD}) by replacing $g_0^2
\to g_0^2 \sigma^{2-D/2}$. The resulting explicit exponent of the scale can 
be combined with the corresponding exponent of an extra external momentum
factor into $(k^2/\sigma)^{D/2-2}$ which was added to the total  
exponent of this ratio on the r.h.s.~of (\ref{GDSEirD}). 
The dimensionless integral therein is written in a form ready to apply the
integration formula (\ref{A1}) of App.~\ref{olliint}. 
The result is $k$-independent as we will see explicitly   
in the next subsection. Therefore, from the non-renormalization of the vertex,
condition {\bf (N1)} which implies $l+m+n=0$, Eq.~(\ref{GDSEirD}) for the
leading infrared contributions to both sides of the ghost DSE entails
that
\begin{equation}
     \epsilon_0+2\delta_0  \, = \,  2 - D/2  \; . \label{irscaleD}
\end{equation}
Thus, the infrared behavior of the running coupling from Eq.~(\ref{alpha_ir}) 
equivalently follows to be of the form,
\begin{equation}
        \alpha(k^2) \to \alpha_c
        \left(\frac{k^2}{\sigma}\right)^{\!\!2-{D}/{2}},  \;\; \mbox{for}
        \quad   k^2 \to  0 \; ,  
\end{equation}  
which is infrared finite for $D=4$. From now on, we therefore parametrise 
the leading infrared exponents of ghost and gluon propagation by the joint
exponent $\kappa $ again, in $D$ dimensions,   
\begin{equation}
    \delta_0 \,\,  = - \kappa   \; , \;\; \epsilon_0 \, = \, 2\kappa +  2 -
    D/2  \; .  
\end{equation}
Infrared enhancement of ghosts follows for exponents
$0<\kappa < (D\!-\!2)/2$ with the
upper bound from the temperedness of local fields. In addition, $ (D\!-\!2)/4
\le \kappa $ for the mass gap in transverse gluon correlations. Together,
this then establishes the Kugo-Ojima confinement criterion
in Landau gauge. Zwanziger's horizon condition furthermore 
excludes equality at the lower bound. We thus expect solutions for
\begin{equation}
 (D-2)/4 < \kappa < (D-2)/2 \; .
\end{equation}
Once this exponent is determined, depending on the infrared exponents of the
ghost-gluon vertex, from Eq.~(\ref{GDSEirD}) we furthermore obtain,
\begin{equation}
    \alpha_c \, = \, \frac{2^{D-2}\pi^{D/2-1}}{N_c\, I_G^{(D)}(\kappa,l,n)} 
  \; ,        \label{alpha_c_G}
\end{equation}
where $I_G^{(D)}(\kappa,l,n)$ is a ratio of $\Gamma$-functions proportional to
the integral in Eq.~(\ref{GDSEirD}) which we determine next.

\section{Infrared Exponent for Ghost-Gluon system, Results}
\label{ieggs}


The same procedure that led to (\ref{GDSEirD}) in the ghost case, can 
be applied to the gluon DSE. In this case, we simply insert the leading 
infrared behavior of the propagators from (\ref{aslZ}), (\ref{aslG}) and the
vertex (\ref{AirAns}) directly into Eq.~(\ref{Zir}). In $D$ dimensions, this
then analogously leads to
\begin{eqnarray}
\label{ZDSEirD}
 \left(\frac{k^2}{\sigma}\right)^{\!\!-\epsilon_0}\hskip -.3cm &=&  
 N_c \, 4\pi\alpha_c \, 
\left(\frac{k^2}{\sigma}\right)^{\!\!2\delta_0+\frac{D}{2}-2}  
 \frac{I_{Z_P}^{(D)}(\kappa,l,n)}{2^D \pi^{D/2}} \;  , \qquad   
\end{eqnarray}
where we used $l+m+n=0$, and again, we conclude the relation for the
exponents in Eq.~(\ref{irscaleD}). A comparison of Eqs.~(\ref{alpha_c_G}) and
(\ref{ZDSEirD}) furthermore tells us that  
\begin{equation}
\label{eq:selfc}
      I_G^{(D)}(\kappa,l,n) \, \stackrel{!}{=} \, I_{Z_P}^{(D)}(\kappa,l,n)
      \; , 
\end{equation}
which determines the values allowed for the exponents $n,l$ and $\kappa $.
The two dimensionless integrals  
$I_G^{(D)}$ and $I_{Z_P}^{(D)}$ for the transverse ghost-loop contribution
are given by,   
\begin{eqnarray}
   \frac{I_{G}^{(D)}(\kappa,l,n)}{2^D \pi^{D/2}} \!&=&\! -\! 
  \int\!\frac{d^Dq}{(2\pi)^D}  
 \left(\frac{1}{q^2}\right)^{\!\!\frac{D}{2}}  \frac{k\mathcal{P}(p)k}{k^2}
    \times \nonumber \\
 &&  \hskip .8cm \left(\frac{q^2}{k^2}\right)^{\!\!\frac{D}{2}-1-\kappa +l}  
     \left(\frac{k^2}{p^2}\right)^{\!\!\frac{D}{2}-1-2\kappa-n} \hskip -.3cm
     , \qquad \label{IGD} \\
   \frac{I_{Z_P}^{(D)}(\kappa,l,n)}{2^D \pi^{D/2}} \!&=&\! 
  \int\!\frac{d^Dq}{(2\pi)^D}  
 \left(\frac{1}{q^2}\right)^{\!\!\frac{D}{2}}
  \frac{q\mathcal{P}(k)q}{(D\!-\!1)k^2} \times \nonumber\\
&& \hskip .8cm   \left(\frac{q^2}{k^2}\right)^{\!\!\frac{D}{2}-1-\kappa -n-l}  
     \left(\frac{k^2}{p^2}\right)^{\!\!1+\kappa -l} \hskip -.3cm . 
                         \label{IZPD}
\end{eqnarray}
To compute these, we first note that
\begin{eqnarray}
\label{proterms}
    \frac{k\mathcal{P}(p)k}{k^2} &=& \frac{1}{2} \left(1+ \frac{q^2}{p^2} 
      +  \frac{q^2}{k^2} \right) - \frac{1}{4} \left(\frac{k^2}{p^2} 
      +  \frac{p^2}{k^2} + \frac{q^4}{k^2p^2}\right)\, ,\nonumber\\
    \frac{q\mathcal{P}(k)q}{k^2} &=& \frac{1}{2} \left(\frac{q^2}{k^2} 
       + \frac{p^2}{k^2} + \frac{p^2q^2}{k^4} \right) - 
     \frac{1}{4} \left(1 + \frac{q^4}{k^4} +  \frac{p^4}{k^4} 
 \right) \, .\nonumber
\end{eqnarray}
With these in Eqs.~(\ref{IGD}) and (\ref{IZPD}), respectively, it is now 
straightforward to apply the formula in Eq.~(\ref{A1}) repeatedly with
suitable substitutions for the exponents $\alpha$ and $\beta$. 
For each of the two integrals this leads to a sum of six ratios of
$\Gamma$-functions with arguments that differ by integer values corresponding
to the above six terms with different powers of the momenta 
in each of the two integrals. After some tedious applications of the
$\Gamma$'s functional identity, these six ratios can be combined into 
a single one, to the effect that,
\begin{widetext}
\begin{eqnarray}
I_{G}^{(D)}(\kappa,l,n) &=& -\frac{D-1}{2}\,
\frac{\Gamma(\frac{D}{2}-\kappa+l)\Gamma(1+2\kappa+n)\Gamma(-\kappa-l-n)}{\Gamma(\frac{D}{2}-2\kappa-n)\Gamma(1+\kappa-l)\Gamma(\frac{D}{2}+1+\kappa+l+n)}
\; , \label{IGDres}\\
I_{Z_P}^{(D)}(\kappa,l,n) &=& \frac{1}{2}\,
\frac{\Gamma(\frac{D}{2}-\kappa+l)\Gamma(1-\frac{D}{2}+2\kappa+n)\Gamma(\frac{D}{2}-\kappa-l-n)}{\Gamma({D}-2\kappa-n)\Gamma(1+\kappa-l)\Gamma(1+\kappa+l+n)}
\; . \label{IZPres} 
\end{eqnarray}
\end{widetext}
Though these quite general results might look complicated at first, in fact, 
they are surprisingly simple. To appreciate this, consider the following 
special case first.

\subsection{Tree-level ghost-gluon vertex}
\label{tlggv}

In this subsection we concentrate on 
the results in the event that the ghost-gluon vertex reduces to
its tree-level form in the infrared, $G^{ir}_\mu = G^0_\mu$. 
This corresponds to $A^{ir} = 1$ and, for $Z_R$, the generalized form with
$B^{ir} = \hat\eta$. We thus set $l=n=0$  
in the general result of Eqs.~(\ref{IGDres}), (\ref{IZPres}), 
and omit the arguments $l$ and $n$ here. In $D=4$ dimensions, this amounts to 
\begin{eqnarray}
I_{G}^{(4)}(\kappa) &=&  -\frac{3}{2}
\frac{\Gamma^2(-\kappa)\Gamma(2\kappa-1)}{\Gamma(-2\kappa)\Gamma(\kappa-1)\Gamma(\kappa+3)}
\; , \label{IG4} \\
I_{Z_P}^{(4)}(\kappa) &=& \frac{1}{2}
\frac{\Gamma^2(-\kappa)\Gamma(2\kappa-1)}{\Gamma^2(\kappa-1)\Gamma(4-2\kappa)}
\; . \label{IZP4}
\end{eqnarray}
The result in Eq.~(\ref{IG4}) then agrees with both versions given for the
tree-level vertex in Ref.~\cite{Atk98}. There, another method was 
employed leading to a rather complicated form which consists of 
sums of confluent hypergeometric functions and which is quite difficult 
to simplify any further. We explain this method and the
connection to ours in App.~\ref{olliint}, {\it c.f.}, Eq.~(\ref{A1})
vs. Eq.~(\ref{A7}), in particular. The two different forms for the result in
Ref.~\cite{Atk98} thereby arose from either choosing the internal gluon or 
the ghost momentum as the integration variable in the loop. Here, both
gives the same in the first place. 
The symmetry $p^2 \leftrightarrow q^2$ 
in the integrand on the l.h.s.~of Eq.~(\ref{A1}) is manifest on the r.h.s by
its explicit invariance under $\alpha \to \alpha' = D/2-\beta$ together
with $\beta\to \beta' = D/2-\alpha $.          

\begin{figure}[t]
\vspace*{.5cm}
\epsfig{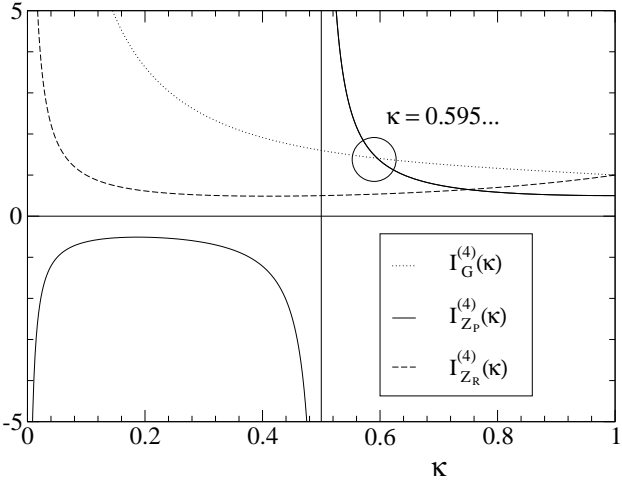}
\caption{\label{tlIs} The Infrared integrals $I_G^{(D)}$,
$I_{Z_P}^{(D)}$ and $I^{(D)}_{Z_R}$ in $D\!=4\!$ dimensions with tree-level 
ghost-gluon vertex, {\it c.f.}, Eqs.~(\ref{IG4}), (\ref{IZP4}) and the
$\eta=1$ case from (\ref{IZR4}), respectively.   
The corresponding value for the exponent $\kappa_1 \approx 0.595 $ is obtained
from the intersection point of $I_G$ and $I_{Z_P}$ as marked by the circle.}
\end{figure}

The selfconsistency condition (\ref{eq:selfc}), for both DSEs to yield the 
same value of the constant $\alpha_c$, is also implemented quite easily in 
Eqs.~(\ref{IG4}) and (\ref{IZP4}). In this case, for the tree-level vertex,  
one derives the condition,
\begin{equation}
      12  (3-2\kappa)  (2\kappa-1)   = (\kappa+2) (\kappa+1) \; .
\end{equation}
This is quadratic in $\kappa$, and the two possibilities are,
\begin{equation}
       \kappa =  \frac{1}{98} \Big(93 \mp \sqrt{1201} \Big)  \; \approx \;
    \big\{
    \,    \underline{0.59535} , \; 1.3026 \, \big\} \; , \label{tlr}
\end{equation}
with one unique root in $0<\kappa_1<1$ which we have underlined in 
(\ref{tlr}). This result was first obtained independently in
Refs.~\cite{Ler01,Zwa01}. The corresponding value of $\alpha_c$ is given by,
\begin{equation}
\label{alpha_c_tl}
             \alpha_c = \frac{4\pi}{N_c I_G^{(4)}(\kappa_1)} \approx 2.9717 \;
             , \;\; \mbox{for} \;\; N_c=3 \; ,
\end{equation}
with $I_G^{(4)}(\kappa_1) = I_{Z_P}^{(4)}(\kappa_1) \approx 1.4096$.
It is smaller by this last factor than 
the value of $\alpha_c = 4\pi/N_c $ derived from $I_G^{(4)}(\kappa) = 1$ for 
$\kappa = 1$ with the tree-level vertex (and also $D\!=\!4$) in
Ref.~\cite{Atk98}. This disagreement is due to the difference between  
the transverse ghost-loop contribution $Z_P$ employed here, 
and the $\mathcal{R}$-contracted $Z_R$ in Ref.~\cite{Atk98}. 
Our calculation for $Z_R$, see below, with $A=1$ and $B=\hat\eta$
for the gerneralized tree-level vertex leads to  
\begin{equation}
\label{IZR4}
I_{Z_R}^{(4)}(\kappa) = 
   \big(4\kappa\!-\!2\big) \Big( 1 - 2 \eta\hat\eta 
       \frac{(2\kappa -3)}{(\kappa -1)}\Big) 
        \,  I_{Z_P}^{(4)}(\kappa) \; .
\end{equation}  
For $\eta=1$ or $0$ this is equivalent to the sum of twelve
confluent hypergeometric functions given as the result for the same integral in
Ref.~\cite{Atk98}. The  $D\!=\!4$ results for the tree-level ghost-gluon 
vertex are summarized in Fig.~\ref{tlIs}. In standard Faddeev-Popov theory
$\eta =1$, and transversality of the ghost loop, $I_{Z_R}^{(4)} =
I_{Z_P}^{(4)}$, required $\kappa = 3/4$.  
In order to tune $Z_R$ for transversality at the selfconsistent value of
the tree-level exponent $\kappa_1$, on the other hand, we would
need $\eta \approx 1.16$ or $-0.16$ which appear to be rather unnatural.  
It is not possible in the ghost-antighost symmetric formulation.

\begin{figure*}
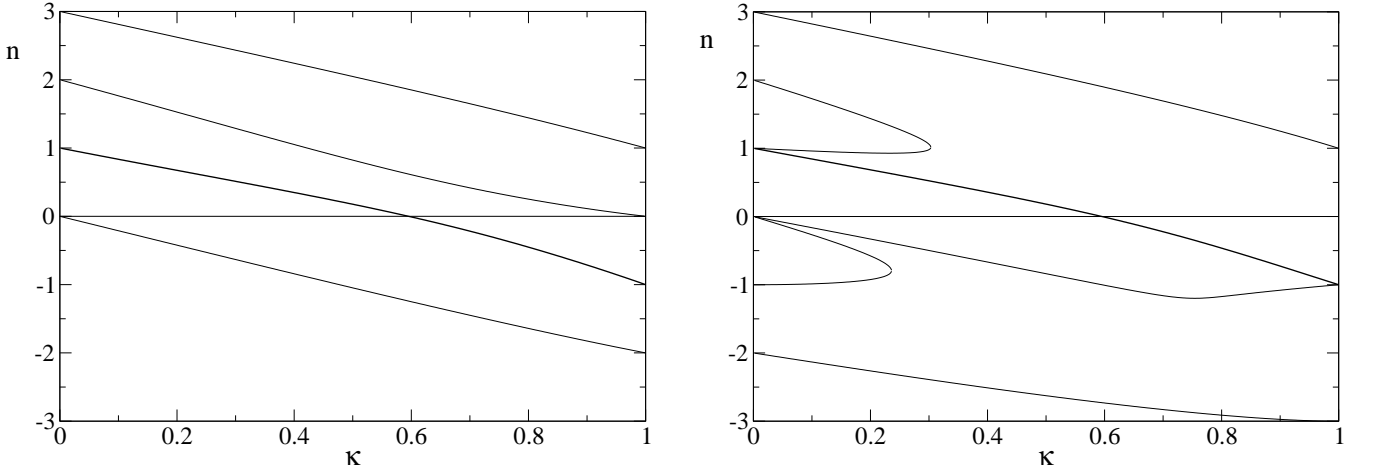

\vspace*{.7cm}
\epsfig{file=my_asymmetric_roots.eps,width=0.48\textwidth}
\hfill
\epsfig{file=my_symmetric_roots.eps,width=0.48\textwidth}
\vspace*{.2cm}
\caption{\label{roots} Roots without symmetrization (left), real roots with
additional symmetrization (right).}
\end{figure*}

Another important difference between 
$I_{Z_P}^{(4)}$ and  $I_{Z_R}^{(4)}$ for the tree-level vertex,  as seen in
Fig.~\ref{tlIs} and Eq.~(\ref{IZR4}), 
is the observation that $I_{Z_P}^{(4)}(\kappa) $ has a pole at $\kappa =1/2$.  
The gluon propagator then necessarily vanishes in the infrared: 
If it was to approach a constant, one had to have $\kappa = 1/2$. 
In this case, however, its constant limit was proportional to 
$1/I_{Z_P}^{(4)}(\kappa) $ which vanishes for $\kappa \to 1/2$.
One thus obtains the strict lower bound $1/2 < \kappa$, 
for the tree-level case. Zwanziger's horizon condition is then satisfied. 
The apparently infrared finite extrapolations from lattice calculations 
are an open question still, however.

\subsection{Infrared transversality}

So much for the tree-level vertex. Before we discuss other, more general
possibilities, in particular, the cases $(i)$ -- $(iii)$ in
Sec.~\ref{subsecVA}, we give a more detailed discussion of the transversality
issue in this subsection.

For the $\mathcal{R}$-contracted infrared contribution of the ghost-loop in
the gluon DSE, Eq.~ (\ref{ZRDSE}), we must specify a form for the second, the
longitudinal $B$-structure of the vertex, in addition.  
The fact that the leading infrared behavior of the propagators should not
depend on the choice of studying $Z_P$ or $Z_R$ can be used to
construct an infrared form of  $B(k^2,p^2,q^2)$ analogous $A^{ir}$ in
(\ref{AirAns}). 

First, we express the integrand in  (\ref{ZRDSE}) for $Z_R$  
in terms of that for $Z_P$ in~(\ref{ZPDSE}) plus a correction term which, for
a given Ansatz $A^{ir}$, is required to vanish (at least upon integration).
For the four terms in the curly brackets in (\ref{ZRDSE}), this leads to
\begin{eqnarray}
  \label{eq:prrel}
  \Big\{ \cdots \Big\} &=&   q\mathcal{P}(k)q \,  A(k^2;q^2,p^2) \,- \,
(D-1) \, \times \\
&& \hskip -1.5cm  \big( \hat\eta k^2 + qk \big) \,
\Bigg( \Big( 1 + \frac{qk}{k^2} \Big) A(k^2;q^2,p^2) \,- \,
       B(k^2;q^2,p^2) \Bigg)\; . \nonumber 
\end{eqnarray}
The first term on the r.h.s herein reproduces  the result for
$I_{Z_P}^{(D)}(\kappa,l,n)$ from  (\ref{IZPD}). With tree-level settings 
$A=1$, $B=\hat\eta$ one readily verifies Eq.~(\ref{IZR4}).

The difference between the infrared integrals, 
$\Delta^{(D)} \equiv  
I_{Z_P}^{(D)} -  I_{Z_R}^{(D)} $, can then be written,
\begin{eqnarray}
\hskip -.4cm    \frac{\Delta^{(D)}}{2^D \pi^{D/2}} \!&=&\! 
  \int\!\frac{d^Dq}{(2\pi)^D}  
 \left(\frac{1}{q^2}\right)^{\!\!\frac{D}{2}}
   \left(\frac{q^2}{k^2}\right)^{\!\!\frac{D}{2}-1-\kappa}  
     \left(\frac{k^2}{p^2}\right)^{\!\!1+\kappa} 
   \hskip -.2cm \times \nonumber\\
&& \hskip -1.8cm  
 \frac{1}{4}\Big( \hat\eta - \eta + \frac{p^2}{k^2}- \frac{q^2}{k^2} \Big)
       \bigg(\big(1 + \frac{p^2}{k^2}- \frac{q^2}{k^2} \big) A^{ir} - 2 B^{ir}
              \bigg) . \;\;   \label{DelD}
\end{eqnarray}
The order of the arguments in $A^{ir}$ and $B^{ir}$ is the same here as that
in (\ref{eq:prrel}) above. The ghost-loop integration projects on terms  
that are overall symmetric in $p^2 \leftrightarrow q^2$. The antisymmetric 
ones vanish upon integration. We use the symmetry {\bf (S1)} of $A$ and 
the decomposition $B = B_+ + B_- $ into (anti)symmetric parts
$B_\pm (x;y,z) = \pm B_\pm (x;z,y)$ for $B$ again, and sort out the symmetric
part of the integrand in (\ref{DelD}). It vanishes, if 
\begin{eqnarray}
(\hat\eta -\eta)\,   A^{ir}(k^2;q^2,p^2)  &=&
                  (\hat\eta -\eta)\,  2B_+^{ir}(k^2;q^2,p^2) 
\; , \;\; \mbox{and} \nonumber \\ 
\frac{p^2 - q^2}{2k^2} \,   A^{ir}(k^2;q^2,p^2)
            &=&  B_-^{ir}(k^2;q^2,p^2) \; .  \label{Bpmir}
\end{eqnarray}
The first condition shows that $ 2B_+^{ir} = A^{ir}$, for $\hat\eta\not=\eta$.
In the symmetric case, $\hat\eta =\eta =1/2$, no such
restriction is implied here, but the same relation is then given by 
{\bf (S2)}, see Sec.~\ref{ggvlg} above. Therefore, Eqs.~(\ref{S2ir}) and 
(\ref{Bmir}) of Sec.~\ref{ggvlg} follow as sufficient conditions for infrared
transversality independent of $\eta$. Without turning them into necessary
conditions this is a rather trivial result.
As we showed in Sec.~\ref{ggvlg}, 
the conditions in~(\ref{S2ir}) and (\ref{Bmir}) imply that the vertex itself
is transverse in the infrared. This is always a possibility to warrant infrared
transversality, however. The necessary condition is $\Delta^{(D)} =0$.  

The point here is to demonstrate that, apart from possible 
accidental cancellations, {\it e.g.}, with tuning  $\eta=1.16$ for the
tree-level case of the last subsection, 
for  general $\eta$, there are really no possibilities
left other than infrared transversality of the vertex itself.
In particular, we should be allowed to choose $\eta$ such as $\eta=1$ for the
standard linear-covariant or $\eta=1/2$ for the ghost-antighost symmetric
case. We might then further say that we are not interested in
contributions to the vertex which themselves vanish upon integration in
Eq.~(\ref{DelD}) for $\Delta^{(D)}\!$. Up to such irrelevant contributions,
which neither contribute to the gluon nor the ghost DSE,    
(\ref{S2ir}) and (\ref{Bmir}) and thus the transversality of the vertex in
the infrared are also necessary conditions for the infrared transversality of
the gluon correlations in Landau gauge independent of $\eta$.

\subsection{Bounds on the infrared exponent}

We now go back to the general results given in Eqs.~(\ref{IGDres}) and
(\ref{IZPres}). 
With these results for the necessary infrared integrals it is little 
effort to explore infrared forms $A^{ir}(k^2;p^2,q^2)$  other than 
$A^{ir} = const.$ for the $A$-structure in the ghost-gluon vertex. 
First, in 4 dimensions, the selfconsistency condition
in Eq.~(\ref{eq:selfc}) for these integrals leads to, 
\begin{eqnarray}
 (l\!+\!n\!+\!\kappa\!-\!1)(l\!+\!n\!+\!\kappa)(l\!+\!n\!+\!\kappa\!+\!1)
 (l\!+\!n\!+\!\kappa\!+\!2)&\stackrel{!}{=}&
       \nonumber \\
&&\hspace{-7cm}-3(n\!+\!2\kappa)(n\!+\!2\kappa\!-\!1)
 (n\!+\!2\kappa\!-\!2)(n\!+\!2\kappa\!-\!3)\;.\qquad \label{gencond}
\end{eqnarray}
This then corresponds to the Ansatz for the vertex 
as given in (\ref{AirAns}), ({\ref{expconst}) with only condition {\bf (N1)} 
being implemented at this stage.   
It is the starting point for the discussion of the three special 
cases introduced in Eqs. (\ref{verti}) -- (\ref{vertiii}) 
of Sec.~\ref{subsecVA}: 

\medskip

\noindent{\bf Case $(i)$.} Here, we simply need to set $n\!+\!l=0$. We then
obtain form (\ref{gencond}) for the possible solutions to
Eq.~(\ref{eq:selfc}) which here reads,
\begin{equation}
  I_G^{(4)}(\kappa,-n,n) \, \stackrel{!}{=} \, I_{Z_P}^{(4)}(\kappa,-n,n)
      \; ,       
\end{equation}
the quartic equation to, {\it e.g.}, determine $n(\kappa)$,
\begin{equation}
 (\kappa\!-\!1)\kappa (\kappa\!+\!1)
 (\kappa\!+\!2) = -3(n\!+\!2\kappa) \cdots (n\!+\!2\kappa\!-\!3)\;.
\end{equation}
The four real roots to this equation are shown for $\kappa $ in $[0,1]$   
in Fig.~\ref{roots} on the left. 

The top branch, with $ 1 \le n \le 3$, leads
to negative $I_G^{(4)} $ and $I_{Z_P}^{(4)}$. Both propagators, $Z(k^2)$ and 
$G(k^2)$, then necessarily had zeros at some finite $k^2$, and it resulted
that $\alpha_c \le 0$ (equality at the bounds $n=3,1$ for $\kappa = 0,1$). We
therefore rule out this solution as being unphysical. 

Also, we are particularly interested in solutions with at most weak
singularities in vertex functions and require therefore $ |n| < 1$. 
This is the case for the next branch with $ 0 < n < 2$ provided 
$0.4222 < \kappa $. This branch will no longer exist after symmetrization
w.r.t. ghost-antighost momenta (see Case $(ii)$ below), however.

For the bottom branch with $ 0 \le n \le -2$, the restriction to weakly
singular vertices, $-1 < n$ leads to $\kappa < 0.4767$. With this solution, 
it is therefore impossible to obtain the mass gap in transverse gluon
correlations, for which $0.5 \le \kappa$. Furthermore, it does not survive the
symmetrization in $(ii)$ either.     

For the branch that includes the tree-level
result, that with $n(\kappa_1)=0$, the critical coupling from
(\ref{alpha_c_G}), with $D=4$ and $N_c=3$, is shown as a function of $\kappa$
in Fig.~\ref{alpha_c_fig} (dashed line). 
Its maximum $\alpha_c^{max} \approx 2.9798 $ occurs
at $\kappa \approx 0.6174$. It is thus slightly larger than the
tree-level value $\alpha_c(\kappa_1) \approx 2.9717$ given in
Eq.~(\ref{alpha_c_tl}).  

\medskip

\noindent{\bf Case $(ii)$.} This case corresponds to a sum of two terms, one
with $m\!=\!0$ and $l\!=\!-n$ as in Case $(i)$ above, and the other with  
$l\!=\!0$ and $m\!=\!-n$. Since the ghost-loop contribution $I_{Z_P}$ 
is manifestly symmetric in $p^2 \leftrightarrow q^2$, this symmetrization
only affects the infrared contribution $I_G$ to the ghost DSE, and we can
therefore write, 
\begin{eqnarray}
\label{ciiIs}
 I_G(\kappa,n) &\equiv& \frac{1}{2} \left( I_G^{(4)}(\kappa,-n,n) +
 I_G^{(4)}(\kappa,0,n) \right) \\
 I_{Z_P}(\kappa,n) &\equiv& I_{Z_P}^{(4)}(\kappa,-n,n) .
\end{eqnarray}
The selfconsistency condition $I_G(\kappa,n)= I_{Z_P}(\kappa,n)$ can now be
used to obtain,
\begin{eqnarray}
\frac{(\kappa\!-\!1)\cdots (\kappa\!+\!2)\; (n\!+\!\kappa\!-\!1)\cdots 
(n\!+\!\kappa\!+\!2)}{(\kappa\!-\!1)\cdots (\kappa\!+\!2) +
(n\!+\!\kappa\!-\!1)\cdots  (n\!+\!\kappa\!+\!2)} &\stackrel{!}{=}& 
       \nonumber \\
&&\hspace{-6.5cm}-\frac{3}{2}(n\!+\!2\kappa)(n\!+\!2\kappa\!-\!1)
 (n\!+\!2\kappa\!-\!2)(n\!+\!2\kappa\!-\!3)\;.\qquad \label{iicond}
\end{eqnarray}
This equation has eight roots $n(\kappa)$ which, in general, come in  
complex conjugate pairs. The real roots for $\kappa$ in $[0,1]$ are shown
in Fig.~\ref{roots} on the right.

One discovers that two out of the originally four branches, from
Case $(i)$ above, are almost
unaffected by the symmetrization employed here: 
These are the unphysical 
one at the top, with  $ 1 \le n \le 3$ and much the same $\alpha_c \le 0$,
and the one connected to the tree-level result, with $n(\kappa_1)=0$ for
$\kappa_1 = (93-\sqrt{1201})/98$. Superimposing both results for the latter,
with and without the symmetrization, the corresponding solutions $n(\kappa)$
turn out to be almost indistinguishable in the whole range $0<\kappa<1$ 
on the scales of our plots. A little bit
more appreciable, but not very significant still, are the differences in 
the corresponding values for $\alpha_c$ as compared to each other in
Fig.~\ref{alpha_c_fig}. For the symmetric solution (solid
line) the maximum occurs at the value $\kappa_1$ for the 
tree-level vertex, $\alpha_c^{max} = \alpha_c(\kappa_1)$, at which both
solutions intersect. As explained in Sec.~\ref{subsecVA}, if we furthermore
require the vertex-structure $A$ to have no infrared divergences associated
with the ghost momenta, we must have $n\le 0$ for the 
ghost-antighost symmetric vertex, in addition. Therefore, we can find 
physical\-ly acceptable solutions in the range 
\begin{equation}
\label{eq:bounds}
      \kappa_1 \le \kappa < 1 \; , \;\; \mbox{and} \;\;\; 
                     0 < \alpha_c(\kappa) \le \alpha_c(\kappa_1) \; ,
\end{equation} 
correspondingly, with the values $\kappa_1 \approx 0.59535$ and 
$\alpha_c(\kappa_1) \approx 2.9717$ of the bounds 
as obtained from Eqs.~(\ref{tlr}) and (\ref{alpha_c_tl})
for the tree-level vertex, respectively.

For completeness we mention that the new branch for the symmterized vertex
$(ii)$ with $-1.2 < n <0$, with possibly interesting solutions 
$-1 \le n $ for $\kappa \le \kappa_1 $, leads to $\alpha_c \le 0 $
throughout. And this is also the case for the bottom branch with $n<-2$.

\begin{figure}[t]
\begin{center}
\epsfig{file=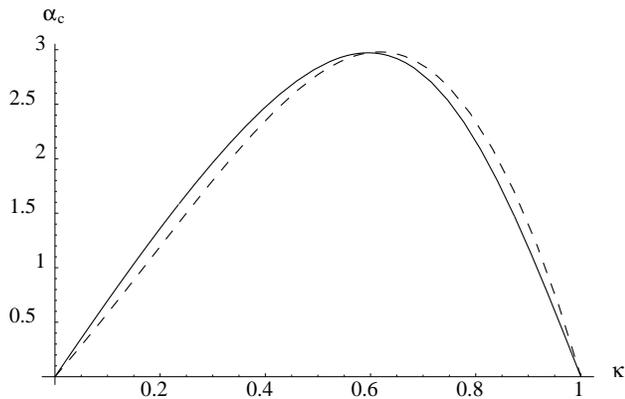,width=0.46\textwidth}
\end{center}
\vspace*{-.4cm}
\caption{\label{alpha_c_fig} The critical value of the running coupling
over the infrared exponent $\kappa$ for Case $(i)$ (dashed) and 
Case $(ii)$ (solid).}
\end{figure}

In the range of particular interest, $1/2 \le \kappa <1$, 
we are thus left with the branch of solutions including $n(\kappa_1)=0$ 
as the only one with $\alpha_c >0$ after the symmetrization in $(ii)$.  
At the same time, it seems quite encouraging that this branch, the only
physically relevant one, is practically unaffected by the symmetrization.

\medskip

\noindent{\bf Case $(iii)$.}~In this example, the infrared behavior of the 
ghost-gluon vertex as given in Eq.~(\ref{vertiii}) is such that it 
again satisfies the conditions {\bf (N1)}, from non-renormalization, 
and {\bf (S1)}, from ghost-antighost symmetry. In the limit
where one ghost momentum vanishes, one now has $A(q^2;q^2,0) = 0$, however.  
We will see that this has no dramatic consequences either, on the physically
interesting solutions found in the range (\ref{eq:bounds}) above.       

Here, by the same arguments as in Case $(ii)$, 
we can now express the leading infrared integrals in both DSEs,
\begin{eqnarray}
\label{ciiiIs}
  I_G(\kappa,n) &\equiv&  I_G^{(4)}(\kappa,-n,n) +
 I_G^{(4)}(\kappa,0,n)  -  I_G^{(4)}(\kappa,0,0)\; , \nonumber\\ 
 I_{Z_P}(\kappa,n) &\equiv& 2I_{Z_P}^{(4)}(\kappa,-n,n) - I_{Z_P}^{(4)}(
 \kappa,0,0) \; .
\end{eqnarray}
Due to the ($l\!=\!n\!=\!0$) contributions herein, which 
arise from the tree-level term (with negative sign) in (\ref{vertiii}),
it is generally no longer possible to derive the solutions to $I_G(\kappa,n)=
I_{Z_P}(\kappa,n)$ as the roots of simple polynomials.
Searching the physically interesting range of parameters numerically, 
starting from the known solution for $n=0$, $\kappa = \kappa_1$, we obtain
the dashed curve for $n(\kappa)$ as compared to the corresponding branch for 
Case $(ii)$ in Fig.~\ref{tlbriii_roots}. Again requiring $n \le 0$ to avoid 
infrared divergent ghost legs, we find that the solutions in the two cases
are remarkably close to each other with $-\kappa < n \le 0$ for $\kappa_1 \le
\kappa < 1$. And again we find that $n \to -\kappa$ in the limit 
$\kappa \to 1$ in which $\alpha_c \to 0$ in all three cases, however.

\begin{figure}[t]
\begin{center}
\epsfig{file=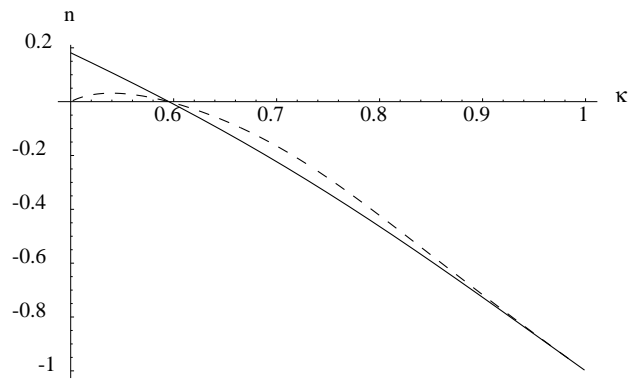,width=0.46\textwidth}
\end{center}
\vspace*{-.4cm}
\caption{\label{tlbriii_roots} 
Solution $n(\kappa)$ for Case $(iii)$ (dashed) compared to 
the tree-level branch of Case $(ii)$ (solid) in the range $1/2 < \kappa <
1$, both with $n(\kappa_1)=0$ at $\kappa_1 = (93-\sqrt{1201})/98$.}
\end{figure}

One might think that another solution with $n=0$ exists for $\kappa=1/2$
and the Case $(iii)$ vertex. 
Since that could have important implications, we note here that this is
actually not the case. We know that $I_{Z_P}$ in (\ref{ciiiIs}) 
reduces to the form for the tree-level vertex at $n=0$, which does not lead
to a solution in 4 dimensions, {\it c.f.}, Fig.~\ref{tlIs}. The Fact that 
the dashed line in Fig.~\ref{tlbriii_roots} appears to approach $n\to 0^+$ for 
$\kappa \to (1/2)^+$ is explained as follows: For sufficiently small
$n=\epsilon $, 
we find from (\ref{ciiiIs}) that 
\begin{equation}
\label{ciiipoles}
   I_{Z_P}(\kappa,\epsilon) \sim \left( \frac{1}{\kappa - (1\!-\!\epsilon)/2} -
\frac{1}{2} \frac{1}{\kappa - 1/2} \right)
 \; , \;\; \mbox{for} \;\; \kappa \to \frac{1}{2} \; , \nonumber
\end{equation}
is dominated by two nearby Poles with opposite signs. 
Therefore, for arbitrarily small but finite $\epsilon >0$ the Pole at
$\kappa =1/2$ in the contribution from the last term, the negative of that in
the tree-level vertex case, will always lead to an intersection, just above 
$\kappa = 1/2$, with $I_G(\kappa,0)$ which approaches a constant 
(corresponding to the value $4\pi/(3I_G(0.5,0)) \approx 2.62$ in
Fig.~\ref{alpha_call_fig}). For $n=\epsilon=0$ on the other hand, both 
Poles coincide and their residues  sum up to that of the tree-level case
which is now positive. And thus, the intersection 
point then disappears. This is confirmed also numerically and demonstrated
in Fig.~\ref{IZiiising}.

Of course, if we relax the condition $n \le 0$, we can have selfconsistent 
solutions also for infrared exponents $\kappa < \kappa_1$, including those
for $\kappa = 1/2 $ in Cases $(i)$ and $(ii)$. At the same time,  
this leads to a singularity in $A(k^2; p^2,q^2)$ as $q^2\to 0$. 

Negative $n$ on the other hand lead to an infrared divergence associated with
the gluon leg. However, as long as $ - n < \kappa $ this is over-compensated
by the gluon propagator attached to that leg, because $Z(k^2) \sim
(k^2)^{2\kappa}$.  
For $n=-\kappa $ an effective massless particle pole would be left 
in a gluon exchange between two vertices, $G_\mu D_{\mu\nu}(k) G_\nu \sim
1/k^2 $.  
In all solutions we report here,  $n \to -\kappa $  for  $\alpha_c \to 0$
(both from above). Therefore, this limit cannot be reached, since then,
at least one of the leading infrared coefficients $d_0$ or $e_0$ in the 
propagators vanishes which contradicts the assumptions, {\it c.f.},
Eqs.~(\ref{aslZ})--(\ref{alpha_ir}). 

No such compensation occurs for divergences associated with ghost legs.
The ghost correlations are themselves infrared enhanced. This infrared
enhancement will persist for the ghost correlations between their vertices, 
if $0 < \kappa - m - l < 1$. Above the upper bound, the infrared divergences 
become too severe for the description in terms of local fields. 
For the cases with the ghost-antighost symmetry {\bf (S1)} of Landau gauge,
we obtain from this restriction the upper bound $ n < (1-\kappa)/2 \le 0.25$
for $\kappa  \le 1/2$. This, however, leaves just 
enough room that it alone does not rule out the solutions with positive $n$
found for $ 1/2 \le \kappa \le \kappa_1$ in case $(ii)$, as seen in
Fig.~\ref{tlbriii_roots}.

\begin{figure}[t]
\begin{center}
\epsfig{file=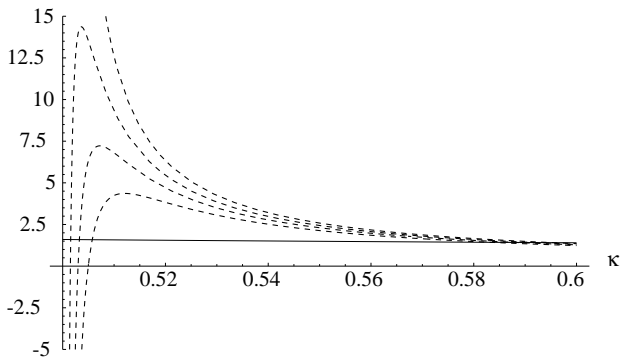,width=0.46\textwidth}
\end{center}
\vspace*{-.4cm}
\caption{\label{IZiiising} 
$I_{Z_P}(\kappa,n)$ (dashed) for several several small values $n= \{0.01,
0.006, 0.003\}$ and $n=0$ over the infrared exponent $\kappa$ in Case $(iii)$.
The intersection with $I_G(\kappa,0)$ (solid) near $\kappa=1/2$ 
disappears for $n=0$.}
\end{figure}


\section{Summary and Conclusions}

The Dyson-Schwinger equations of standard Faddeev-Popov theory
in Landau gauge, when supplemented by additional boundary conditions, can be
derived as an approximation to the time-independent diffusion equation of
stochastic quantization which is valid non-perturbatively \cite{Zwa01}. 
The non-conservative part of the drift force that is neglected in this
approximation cannot be described by local interactions. The effect of 
this part will have to be investigated in future. It may well be
responsible for the kind of ``Gribov noise'' observed in lattice
calculations.  

\begin{figure}[t]
\begin{center}
\epsfig{file=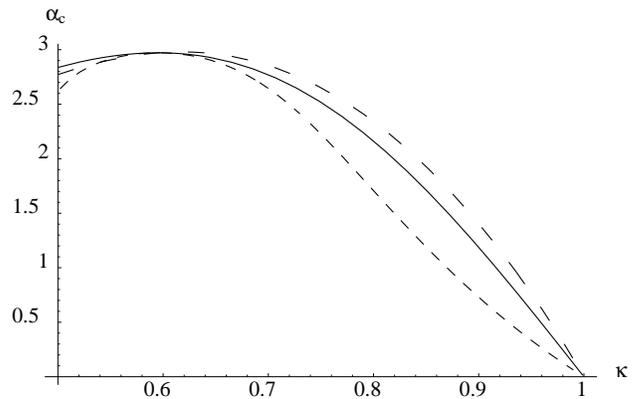,width=0.46\textwidth}
\end{center}
\vspace*{-.4cm}
\caption{\label{alpha_call_fig} The value of $\alpha_c$  
in the range $1/2 < \kappa < 1$ 
for all three Cases, $(i)$ (long dashed), 
$(ii)$ (solid), and $(iii)$ (short dashed).}
\end{figure}

Here, we studied a slightly more general definition of the Landau gauge
as limit of a wider class including non-linear covariant gauges \cite{Bau82}. 
This limit is controlled by an additional free parameter $\eta$ 
in the tree-level vertex (with $\eta = 1$ in Faddeev-Popov theory). In
particular, we find that non-renormalization of the vertex in a symmetric
subtraction scheme and infrared transversality of the gluon propagator in
Landau gauge can only go together with the manifestly ghost-antighost
symmetric choice $\eta = 1/2$. In the light of the recent progress connecting
the linear-covariant gauge with time-independent stochastic quantization, 
the ghost-antighost symmetric Curci-Ferrari gauges might therefore also
deserve to be reconsidered for similar connections. 
 
Optimistically assuming that perfect sense can be made of Dyson-Schwinger
equations non-perturbatively some day, we studied the infrared critical
exponent and coupling for gluon and ghost propagation in Landau gauge 
in quite some generality. We gave two reasons for assuming ghost dominance,  
the Kugo-Ojima criterion for confinement and the horizon condition to
restrict the measure to the first Gribov region, and implemented this as 
a boundary condition in our infrared asymptotic discussion of the
DSE solutions. Central to an understanding of the infrared exponents
for gluon and ghost propagation in Landau gauge then is a knowlegde of their
vertex. Assuming it has a regular infrared limit, we obtain  
$\kappa \simeq 0.595 $. For the ghost-antighost symmetric vertices, this
value maximizes the critical coupling $\alpha_c(\kappa)$, yielding
$\alpha_c^{max} \approx 2.97 $, as summarized once more 
for $\kappa $ between $0.5$ and $1$ in Fig.~\ref{alpha_call_fig}. For larger
$\kappa $ the vertex acquires an infrared singularity in the gluon momentum,
smaller ones imply infrared singular ghost legs. 

Quite encouragingly, numerical solutions to truncated Dyson-Schwinger
equations have recently been obtained with the infrared behavior 
of the regular vertex, as obtained here,
for the whole momentum range up to the perturbative ultraviolet
regime and without one-dimensional approximation in Ref.~\cite{Fis02}.  

An important detail in all our considerations is the non-renormalization of
the ghost-gluon vertex in Landau gauge. Derived from standard Slavnov-Taylor
identities it is one of the arguments that hold at all orders of
perturbation theory. That it is also true non-perturbatively, however, is
another additional assumption. It is therefore quite important and
interesting that this has been assessed and verified within
the numerical errors in calculations using Landau gauge on the lattice
\cite{Lan02}. Calculating both propagators simultaneously, this study
furthermore appears to confirm a unique exponent for the combined 
infrared behavior of gluons and ghosts for the first time in a lattice
calculation implementing the Landau gauge condition.  Also, for $SU(2)$ this
study reports preliminary values of $\alpha_c$ that are fully consistent with
the results obtained here somewhere near the maximum $\alpha_c^{max} \simeq
(4\pi/N_c)\,  0.71 \approx 4.5 $ for $N_c=2$. It will be very interesting to 
see the final errors, so that we will be in the fortunate position to
restrict further the range of both, $\alpha_c$ and  $\kappa $. 
At the moment, the combined evidence seems to indicate that the result will
be somewhere in the range around $\kappa= 0.5$ and the maximum near 
$\kappa =0.6$. Unfortunately, with this conclusion 
the question about an infrared vanishing versus finite gluon propagator
must therefore remain open, for the time being.


\vspace*{-.2cm}

\begin{acknowledgments}
We gratefully acknowledge discussions and communications 
with R.~Alkofer, C.~Fischer, K.~Langfeld, F.~Lenz,
J.~M.~Pawlowski, A.~G.~Williams, and D.~Zwanziger.
\end{acknowledgments}

\medskip 

\centerline{\small\bf Notes added}

\medskip 

P. Petreczky kindly reminded us of the lattice Landau gauge results
for the 3-dimensional gluon propagator of Refs.~\cite{Cuc01a,Cuc01b}. 
We gratefully acknowledge communications with him on their results.
 
From communications with A. Davydychev, we learned that we might
have inadvertently made the impression to consider formula (\ref{A1}) 
as new in any sense. This is not at all the case. The 2-line
derivation in (B2), (B3) below is given for the convenience to the reader.

He furthermore points out that relation (\ref{A13}) follows from 
Eq. (11) listed on page 534 in the tables of Ref.~\cite{Pru90}.
We gratefully acknowledge this information.


\appendix

\section{G\lowercase{host} RG \lowercase{equation revisited}}
\label{corrRG}

We repeat the renormalization group (RG) analysis of Ref.~\cite{Wat01} 
for the ghost propagator with some minor corrections. These corrections
do not affect the main conclusions of Ref.~\cite{Wat01} as far as we judge.
The correct versions, in particular, of Eqs. (8) to (11) in Ref.~\cite{Wat01},
are necessary, however, in order to establish the equivalence of their 
asymptotic infrared expansion and the one adopted via Eq.~(\ref{asG})
herein, which is a minor variation of the expansion techniques developed
previously \cite{Atk81,Atk82,Hau96a,Hau96b,Hau96,Hau98,Sme98}.  

First, recall the RG equation for the ghost propagator $G(k^2) \equiv
G(k^2,\mu^2)$, in this form also given in (6) of \cite{Wat01},
\begin{equation}
\label{RGG}
 \left( \mu\frac{\partial}{\partial \mu} + \beta(g) \frac{\partial}{\partial
 g} - 2 \gamma_G(g) \right) G(k^2,\mu^2) = 0 \;. 
\end{equation}
The formal solution to this equation is given by Eq.~(\ref{RGsG}).
With the infrared Ansatz in the form of Eq.~(5) in \cite{Wat01},
\begin{equation}
\label{irWat}
    G(k^2,\mu^2) \simeq    \sum_n^N
   d'_n(g)   \left(\frac{k^2}{\mu^2}\right)^{\delta_n}   \; ,
\end{equation}
here denoting the coefficients  of \cite{Wat01} as the primed ones, $d_n'$, 
to distinguish from those in (\ref{asG}), we first 
obtain,
\begin{equation}
\label{Wat:8}
   \beta(g) \left(\frac{\partial d_n'}{\partial g} +  d'_n\ln(k^2/\mu^2) 
      \frac{\partial \delta_n}{\partial g} \right) - 2 d'_n\big(\delta_n
      +\gamma_G(g)\big) = 0 \, , 
\end{equation}
at variance with Eq.~(8) of Ref.~\cite{Wat01} in two minor 
ways (by the factor of 2 and the sign of the $\gamma_G(g)$-term). 
Nevertheless, with their conclusion that therefore ${\partial
\delta_n}/{\partial g} = 0$, we find for the coefficients,
\begin{equation}
     \frac{\partial d'_n}{\partial g} = \frac{2\big(\delta_n +
     \gamma_G(g)\big) }{\beta(g)}\,  d'_n \; ,
\label{RGdpr}
\end{equation}
the general solution of which takes the form,
\begin{equation}
     d'_n(g)  \propto \exp \int^g  \frac{2\big(\delta_n +
     \gamma_G(l)\big)}{\beta(l)} dl \; . \label{dprs}
\end{equation}
This, however, is incompatible with Eq.~(9) of Ref.~\cite{Wat01},
\begin{equation}
     d'_n(g)  = \mbox{const.} 
  \;  g^{- \frac{2(\delta_n + \gamma_G)}{2\gamma_G + \gamma_A}} 
\label{Watfor}. 
\end{equation}
In particular, since we just noted that the exponents $\delta_n$ are
$g$-independent, they are either zero or one would need 
$\beta(g) = - \mbox{const.} \times g $ to obtain (\ref{Watfor})
from (\ref{dprs}). By virtue of Eq.~(\ref{andim}),  
$\beta(g) = -g (2\gamma_G + \gamma_A) \sim -g $, this would 
imply that $2\gamma_G + \gamma_A$ is $g$-independent. If we
would then conclude in addition that both, $\gamma_G$ and $\gamma_A$, are
$g$-independent, only then we would obtain (\ref{Watfor}) from (\ref{dprs}).
 
Note that such a behavior, $\beta(g) \sim -g $, though in
principle possible in the infrared, would not lead to a fixed point and thus
contradict the other results of \cite{Wat01} as we discussed in
Sec.~\ref{iarg} above. Not restricted to such a specific behavior, here we 
go back to the general form of the $d'_n$ in Eq.~(\ref{dprs}). First, 
remember that $g= g_0 $ for $\mu^2 = \sigma$. In this case, the exponential
factor in our expansion (\ref{asG}) becomes unity, and (\ref{asG}) and
(\ref{irWat}) agree. Thus, $d_n = d'_n(g_0)$, and we can 
split the solution  to (\ref{RGdpr}) for $d_n'$ with this initial 
condition into factors as follows, 
\begin{equation}
\label{dpfac}
  \hskip -.2cm    
     d_n'(g) = d_n \exp\big\{
     2\delta_n\int^g_{g_0} \frac{dl}{\beta(l)} \big\}  
    \, \exp\big\{2\int^g_{g_0} \frac{\gamma_G(l)}{\beta(l)} dl \big\} \,  .
\end{equation}
The first exponential factor herein, with (\ref{smug}), is equal to 
$(\mu^2/\sigma)^{\delta_n}$, and can be used to replace $\mu^2 \to \sigma $
in (\ref{irWat}). The last exponential in (\ref{dpfac}) is the same
overall factor determining the $(g,\mu)$-dependence of the ghost propagator 
as in (\ref{asG}). Substituting (\ref{dpfac}) into  the
expansion (\ref{irWat}) of Ref.~\cite{Wat01}, one obtains, 
\begin{equation}
\label{irme}
  G(k^2,\mu^2) \simeq      \exp\big\{2\int^g_{g_0}
                  \frac{\gamma_G(l)}{\beta(l)} dl \big\} \,    \sum_n^N
                  d_n  \left(\frac{k^2}{\sigma}\right)^{\delta_n} \!, 
\end{equation}
which agrees with the RG invariant expansion of (\ref{asG}).

\section{T\lowercase{wo ways to do the} $D$-\lowercase{dimensional integrals}}
\label{olliint}

The basic formula we employ for the infrared analysis in Sec. \ref{riia} 
involves $D$-dimensional integrals of the following form
which converge for $\mbox{Re}(\alpha) > 0$, $\mbox{Re}(\beta -\alpha) > 0$, 
$\mbox{Re}(\beta) < D/2$,
\begin{eqnarray}
 && \int  \frac{d^Dq}{(2\pi)^D}  \,   \left(\frac{1}{q^2}\right)^{D/2} 
\,   \left(\frac{q^2}{k^2}\right)^\alpha  \,
\left(\frac{k^2}{p^2}\right)^\beta \, = \label{A1} \\
 && \hskip 1cm  \frac{1}{2^D \pi^{D/2}} \,
\frac{\Gamma(\alpha) \Gamma(D/2 - \beta)\Gamma(\beta
-\alpha)}{\Gamma(\beta)\Gamma(D/2-\alpha)\Gamma(D/2 + \alpha -\beta)} \; ,
  \nonumber
\end{eqnarray}  
where $p=k\pm q$ and an explicit factor $(k^2)^{\beta -\alpha}$ 
was introduced to render the integral dimensionless. 
This is a textbook formula, {\it c.f.}, Eq.~(2.5.178) in \cite{Muta}.
For a simple derivation with the general exponents \cite{Sch00p} one 
observes that the l.h.s.~of (\ref{A1}) is a convolution
integral which reduces to an ordinary product upon Fourier
transformation. Using  
\begin{eqnarray}
 \frac{1}{(q^2)^\gamma} =  
 \frac{\Gamma(D/2-\gamma)}{4^{\gamma} \pi^{D/2} \Gamma(\gamma)} 
\int  
     d^D\!x  \; (x^2)^{\gamma-D/2} \, e^{-iqx} \; , 
\label{A2}
\end{eqnarray} 
for the two factors in the convolution (with the power $\gamma$ 
given by $D/2-\alpha$ and $\beta$, respectively), one thus obtains,  
\begin{eqnarray}
 && \int  \frac{d^Dq}{(2\pi)^D}  \,   \left(\frac{1}{q^2}\right)^{D/2} 
\,   \left(\frac{q^2}{k^2}\right)^\alpha  \,
\left(\frac{k^2}{p^2}\right)^\beta \, = \label{A3} \\
 && \hskip .1cm  \frac{4^{\alpha-\beta}}{(2\pi)^D} \,
\frac{\Gamma(\alpha) \Gamma(D/2 - \beta)}{\Gamma(\beta)\Gamma(D/2-\alpha)} 
 \int \! \frac{d^D\!x}{(x^2)^{D/2}} \,
 (k^2 x^2)^{\beta - \alpha} \,\, e^{-ikx} \; .
  \nonumber
\end{eqnarray}  
Eq. (\ref{A1}) then follows from a further application of the Fourier
transform (\ref{A2}) herein, now with $\gamma = \beta - \alpha$.

Alternatively, we can do the integral in (\ref{A1}) which we denote
henceforth by $f_D(\alpha,\beta)$ in a straightforward though less elegant way 
by first performing all but one of the angles of the polar coordinates in
$D$-dimensional momentum space,
\begin{eqnarray} 
f_D(\alpha,\beta) &=& \frac{K(D)}{(2\pi)^D} \frac{D}{2} \int \frac{dq^2}{q^2}
\,   \left(\frac{q^2}{k^2}\right)^\alpha  \, \times  \label{A4} \\
      && \hskip -1cm \frac{1}{B((D\!-\!1)/2,1/2)} \int_{-1}^1 \!dz  \, (1-z^2)^{(D-3)/2}  \,
\left(\frac{k^2}{p^2}\right)^\beta \; , \nonumber
\end{eqnarray}
where $K(D) = 2\pi^{D/2}/(D\Gamma(D/2)) $ is the volume of the
$D$-dimensional unit ball, $p^2 = q^2 + k^2 - 2kq z$, and the Euler Beta
function is given by
\begin{eqnarray}
B((D\!-\!1)/2,1/2) &=& \int^1_0 dt \,  t^{-1/2} (1-t)^{(D-3)/2} \nonumber \\
&=&         \int_{-1}^1 \!dz  \, (1-z^2)^{(D-3)/2}  \; . 
\end{eqnarray}
For the azimuthal integration in (\ref{A4}), the formula 2. in {\bf 3.665} of
Ref.~\cite{Gra94} can be used to obtain,
\begin{widetext}
\begin{eqnarray} 
f_D(\alpha,\beta) &=& \frac{K(D)}{(2\pi)^D} \frac{D}{2} \left(
\int_0^{k^2}  \frac{dq^2}{q^2}
\,   \left(\frac{q^2}{k^2}\right)^\alpha  \!\!
        \ _2F_1\Big(\beta,\beta\!-\!\frac{D}{2}\!+\!1;
\frac{D}{2};\frac{q^2}{k^2}\Big) 
 +    \int_{k^2}^\infty  \frac{dq^2}{q^2}
\,   \left(\frac{q^2}{k^2}\right)^{\alpha-\beta}   
     \!\!\!\!\!\!\!\!  \   _2F_1\Big(\beta,\beta\!-\!\frac{D}{2}\!+\!1;
\frac{D}{2};\frac{k^2}{q^2}\Big) \right)   \nonumber \\
 &=& \frac{1}{2^D \pi^{D/2}\Gamma(D/2)} \left(
\int_0^1  \frac{dx}{x}
\,   \Big( x^\alpha + x^{\beta-\alpha} \Big)   
\ _2F_1\Big(\beta,\beta\!-\!\frac{D}{2}\!+\!1;\frac{D}{2};x\Big) 
         \right) \label{A6} \\ 
&& \hskip -1.5cm =   
\frac{1}{2^D \pi^{D/2}\Gamma(D/2)} \left( \frac{1}{\alpha} 
\ _3F_2\Big(\beta,\beta\!-\!\frac{D}{2}\!+\!1,\alpha;
         \frac{D}{2},\alpha\!+\!1;1\Big)
+ \frac{1}{\beta-\alpha} 
            \ _3F_2\Big(\beta,\beta\!-\!\frac{D}{2}\!+\!1,\beta\!-\!\alpha;
\frac{D}{2}, \beta\!-\!\alpha\!+\!1;1\Big)  \right)  , \label{A7}
\end{eqnarray}
\end{widetext}
where for the last step the following integration formula for the generalized
hypergeometric functions led to the final result in (\ref{A7}), 
 \begin{eqnarray} 
     \int_0^1  dx \, x^{\gamma}  \, \  _pF_q({\bf a};{\bf b};x)  &=&
\\
&& \hskip -2cm     \frac{1}{\gamma +1}  \  _{p+1}F_{q+1}(\{{\bf
a},\gamma\!+\!1\};\{{\bf b},\gamma\!+\!2\};1) \; , \nonumber 
\end{eqnarray}
which is most easily derived for $\gamma > -1$ (see, {\it e.g.}, the
appendices of Refs.~\cite{Atk98,Ler01}) from the power series expansion for
the generalized hypergeometric functions,
\begin{eqnarray} 
    \  _pF_q({\bf a};{\bf b};z)  
= \sum_{n=0}^\infty  \frac{(a_1)_n \cdots (a_p)_n}{(b_1)_n
\cdots (b_q)_n} \, \frac{z^n}{n!} \; , 
\end{eqnarray}
by noting the relation,
\begin{equation} 
        \frac{1}{\gamma+n+1} \, = \,  \frac{1}{\gamma+1} \, 
        \frac{(\gamma+1)_n}{(\gamma+2)_n} \; , 
\end{equation}
for the Pochhammer symbols,
\begin{equation} 
       (a)_n  = \frac{\Gamma(a+n)}{\Gamma(a)} \; .
\end{equation}
For other general properties and a variety of relations amongst the 
different hypergeometric functions, {\it e.g.}, see
Refs. \cite{Erd53,Abr74,Wen00,Bue01}. 
A well-known one for example expresses the Gauss series as a ratio of
Gamma functions,  
\begin{equation} 
              \ _2F_1(a,b;c;1)\,= \,
 \frac{\Gamma(c)\Gamma(c-a-b)}{\Gamma(c-a)\Gamma(c-b)} \; . \label{A12}
\end{equation}
Many more relations of this kind, including less known ones, 
are listed in the tables on hypergeometric functions of Ref.~\cite{Pru90}.
We originally thought it might be interesting to note that the two ways to
calculate $f_D(\alpha,\beta)$ leading to the r.h.s. of Eqs. (\ref{A1}) and
(\ref{A7}), respectively, allow to devise some of these additional
relations. 
For example, by simple renamings, a comparison of the r.h.s.~in
(\ref{A1}) and (\ref{A7}) yields,
\begin{widetext}
\begin{eqnarray} 
           \frac{1}{a}  \, \ _3F_2(a,a\!+\!b,a \!+\!b\!-\!c\!+\!1;c,a
           \!+\!1;1) \,+\,  \frac{1}{b}  \, \ _3F_2(b,a\!+\!b,a
           \!+\!b\!-\!c\!+\!1;c,b \!+\!1;1) &=&
\frac{\Gamma(a)\Gamma(b)}{\Gamma(a+b)} \,  
 \frac{\Gamma(c)\Gamma(c-a-b)}{\Gamma(c-a)\Gamma(c-b)} \nonumber\\
   &=&  B(a,b) \, \ _2F_1(a,b;c;1)  \; . \qquad \label{A13}
\end{eqnarray}
\end{widetext}
\vspace{-2cm}
This formula follows with replacing $b\to a+b$, $1-c\to c-a-b$, and $d\to c$
upon rearrangement from Eq.~(11) on p. 534 in~\cite{Pru90}, and our
presentation here seems obsolete now \footnote{This was not known to us 
originally. We thank A. Davydychev for bringing it to our attention, see the
notes added.}. At least, the equivalence of the results from the infrared
analysis of DSEs in Ref.~\cite{Atk98}, to the expressions in Eqs.~(\ref{IG4})
and (\ref{IZR4}) with $\eta=0$ for the tree-level vertex case of
Sec.~\ref{tlggv} is explicitly established in this way.    
The second procedure to calculate integrals such as
$f_D(\alpha,\beta)$ was thereby used in Ref.~\cite{Atk98}.
Each of the results therein are readily expressed in terms of
one simple ratio of Gamma functions when using the relations
presented in this appendix.
Though equivalent to Eq.~(\ref{A7}) of course, 
use of Eq.~(\ref{A1}) thereby is far more convenient for all practical
purposes. 




\end{document}